\documentclass[12pt, a4paper]{article}
\usepackage{arxiv}
\usepackage[utf8]{inputenc} 
\usepackage[T1]{fontenc}    
\usepackage[numbers]{natbib}
\usepackage{hyperref}       
\usepackage{url}            
\usepackage{booktabs}       
\usepackage{amsfonts}       
\usepackage{amsmath}
\usepackage{nicefrac}       
\usepackage{microtype}      
\usepackage{lipsum}
\usepackage{bm}
\usepackage{soul}
\usepackage{setspace}
\usepackage[title]{appendix}

\usepackage{graphicx}
\usepackage{subcaption}
\usepackage[inline]{enumitem}
\usepackage[symbol]{footmisc}

\usepackage[dvipsnames]{xcolor}

\DeclareMathOperator*{\argmax}{arg\,max}

\title{Where to Drill Next? A Dual-Weighted Approach to Adaptive Optimal Design of Groundwater Surveys}

\author{
  Mikkel B.~Lykkegaard\thanks{Corresponding author.} \\
  Centre for Water Systems and\\
  Institute for Data Science and AI\\
  University of Exeter\\
  Exeter, EX44QF \\
  \texttt{m.lykkegaard@exeter.ac.uk} \\
   \And
 Tim J.~Dodwell \\
  The Alan Turing Institute and\\
  Institute for Data Science and AI\\
  University of Exeter\\
  Exeter, EX44QF \\
  \texttt{tdodwell@turing.ac.uk} \\
}

\begin{document}
\maketitle
\setstretch{1.3}
\begin{abstract}
We present a novel approach to adaptive optimal design of groundwater surveys -- a methodology for choosing the location of the next monitoring well. Our dual-weighted approach borrows ideas from Bayesian Optimisation and goal-oriented error estimation to propose the next monitoring well, given that some data is already available from existing wells. Our method is distinct from other optimal design strategies in that it does not rely on Fisher Information and it instead directly exploits the posterior uncertainty and the expected solution to a dual (or \textit{adjoint}) problem to construct an acquisition function that optimally reduces the uncertainty in the model as a whole and some engineering quantity of interest in particular. We demonstrate our approach in the context of 2D groundwater flow example and show that the dual-weighted approach outperforms the baseline approach with respect to reducing the error in the posterior estimate of the quantity of interest.
\end{abstract}


\section{Introduction}
In this paper, we present a novel approach to optimally choosing the location of the next monitoring well when conducting a groundwater survey. Establishing a monitoring well is generally costly, and depends on the specific geological context and the required penetration depth, and choosing the most informative location for each well is a critical task when designing a groundwater survey. Groundwater surveying and modelling are intrinsically imbued with uncertainty and solutions and predictions are without exception non-unique \citep{anderson_applied_2015}. Hence, in this paper we assume the perspective that a useful sampling location is one that most significantly reduces the uncertainty in the solution, while simultaneously having a substantial influence on some quantity of interest (QoI). While multiple non-invasive and relatively inexpensive methods for groundwater surveying exist \citep{loke_recent_2013, saey_combining_2015, nielsen_review_2017, auken_ttem_2019}, these methods all involve solving an inverse problem to reconstruct the hydraulic head, which introduces an additional layer of uncertainty. Hence, in this work, we focus on the problem of determining aquifer characteristics from direct point measurements of hydraulic head and flux from monitoring wells, and how to optimally choose the locations of such wells, given existing data. While the method is here contextualised within this particular problem, it can easily be generalised to any setting where a continuous function and a derived QoI are approximated with point measurements.

In the ``classic'' theory of optimal design, we often distinguish between optimality criteria that minimise the estimated parameter variances (e.g. $A-$, $D-$ and $E-$optimality) and those that minimise the prediction variance (e.g. $G-$, $V-$ and $I-$optimality) \citep{pukelsheim_optimal_2006, myers_response_2016}. Since in this study we are primarily concerned with the prediction variance, the method presented here belongs in the latter category. In this context, our method can broadly be considered $G-$optimal, since our vanilla acquisition function targets the location of the highest posterior dispersion \citep[see e.g.][]{myers_response_2016}. However, rather than iteratively searching for a design that maximises an optimality criterion, we directly utilise a posterior dispersion estimate to construct an acquisition function. We remark that while there are some abstract parallels between the method presented here and classic optimal design, our method is probably better understood in the context of Bayesian Optimisation, as discussed later. Additionally, the classic optimal design approach is typically centered around the problem of choosing an experimental design that is optimal with respect to an optimality criterion, \textit{before taking any measurements}. In this paper, we take an adaptive approach and assume that some measurements are already available, and we want to propose optimal \textit{new} sampling locations, given the data we already have. How the initial measurement locations are optimally chosen is beyond the scope of this paper, but we refer to e.g. \citet{cox_theory_2000, pukelsheim_optimal_2006, myers_response_2016} for an extensive overview of optimal design of experiments. We remark that our dual-weighted method could in theory be employed to choose initial measurement locations, but in that case the dispersion of the solution would be constrained only by the prior distribution of parameters and the constraints imposed by the constitutive equations. In this case, the method presented herein may be used in conjunction with some space-filling design strategy or using local penalisation functions as described in Section \ref{sec:dual_weighted}. However, either of these workarounds would require an informed prior to work well.

We recycle the notion from classic optimal design that the information gain is driven by minimising the dispersion of a target distribution \citep{lindley_measure_1956}. However, rather than integrating out all possible measurements and model parameters to find the utility of a given design, we take a simpler approach. Namely, we use a Monte Carlo estimate of the (current) posterior dispersion of the solution to a Partial Differential Equation (PDE) (or some appropriate function thereof) as an acquisition function. The underlying rationale being that if we wish to know more about the distribution of our solution, the most useful place to take a new sample is at the point of the highest posterior uncertainty.

In this context, our Vanilla approach (see Section \ref{sec:vanilla}) is not dissimilar to the maximum entropy approach to the optimal sensor placement problem \citep{shewry_maximum_1987}, where sensors are added at the point of the highest uncertainty of some probabilistic function that is fitted to current sensor measurements, for example a Gaussian Process (GP) emulator. While this strategy will typically place many sensors at the boundaries of the sampling space in the context of adaptive GP fitting \citep{mohammadi2021crossvalidation}, this is not necessarily the case when targeting the uncertainty of the solution to a PDE, since that will be constrained by boundary conditions. The sensor placement problem has been studied extensively in the context of GP emulators, and multiple improvements to the maximum entropy approach have been made (see e.g. \citet{krause_near-optimal_2008, beck_sequential_2016, mohammadi2021crossvalidation}). However, since our objective is to minimise the uncertainty of a PDE-derived QoI, and not a GP emulator, many of the recent developments are not immediately applicable, since they are tailored for use with a GP emulator. Hence, the Vanilla approach presented herein can be considered a reformulation of the original maximum entropy approach, particularly tailored for the (probabilistic) solution of a PDE.

Our method (see Section \ref{sec:ada_design}) borrows ideas from other fields, not obviously related to classic optimal design. First, our adaptive optimal design approach is formulated in terms of an acquisition function, a term typically associated with Bayesian Optimisation (BO, \citet{mockus_bayesian_1989, frazier_tutorial_2018}). Moreover, our approach uses ideas from both prior-guided BO \citep{souza_bayesian_2021} and batch BO \citep{pmlr-v51-gonzalez16a}, the similarities with which are discussed in Section \ref{sec:remarks}. While in the context of BO, the aim is to find the maximum or minimum of some function that is expensive to evaluate, our objective is to simply reduce the uncertainty of our model predictions. Hence, our vanilla acquisition function addresses solely the uncertainty of some target function, and not the function value itself. Second, our approach is inspired by the goal-oriented error-estimation used in mesh-adaptation for PDEs \citep{prudhomme_goal-oriented_1999, oden_goal-oriented_2001}, where the intention is to refine a mesh locally and parsimoniously to reduce the simulation error with respect to some QoI using an influence function that is the solution to an adjoint PDE. This approach, however, is most useful for forward problems, where the domain and coefficients are well-known, and the groundwater flow problem is typically not of this kind. Instead, we use the same approach of computing an influence function with respect to the QoI to determine, not where the mesh should be refined, but from where we need more data.

The idea of exploiting the adjoint or \textit{dual} problem to minimise the posterior uncertainty with respect to a QoI was first explored by \citet{attia_goal-oriented_2018} in a similar context as our model problem. However, there are several crucial differences between their approach and the one presented in this paper. First, their method is set in the ``classic'' optimal design context, where a number of sampling locations are determined before taking any measurements, based on the maximising the expected information gain according to some criterion derived from the Fisher Information matrix. Second, since only a finite number of designs can be explored this way, the prospective sampling locations are fixed to a relatively coarse grid. Finally, the approach described in \citet{attia_goal-oriented_2018} requires the adjoint operator to be linear -- an assumption which is suitable for only a subset of QoIs.

We employ Markov Chain Monte Carlo (MCMC) techniques (see Section \ref{sec:bayes_inv}) to generate samples from the posterior distribution of the model parameters given the data $\pi(\theta \vert \mathbf{d})$, where the model parameters $(\theta)$ in this case describe hydraulic conductivity and the data ($\mathbf{d}$) are point measurements of hydraulic head and flux (see Section \ref{sec:gw_flow}). Even if the model parameters themselves are of secondary interest to a given problem, we can use the MCMC samples to construct Monte Carlo estimates of any parameter-derived quantity or function, such as the hydraulic flux across a boundary, or the peak concentration of a contaminant at a well. Additionally, unlike traditional inversion techniques, MCMC allows for rigorously quantifying the uncertainty of the inverse problem, which is useful in the context of engineering decision support systems, in particular risk assessment studies. We believe that there are many unexploited application opportunities tangential to the study of Bayesian posteriors and demonstrate, in this paper, one such application.

\begin{figure}[htbp]
  \centering
  \includegraphics[width=1.0\textwidth]{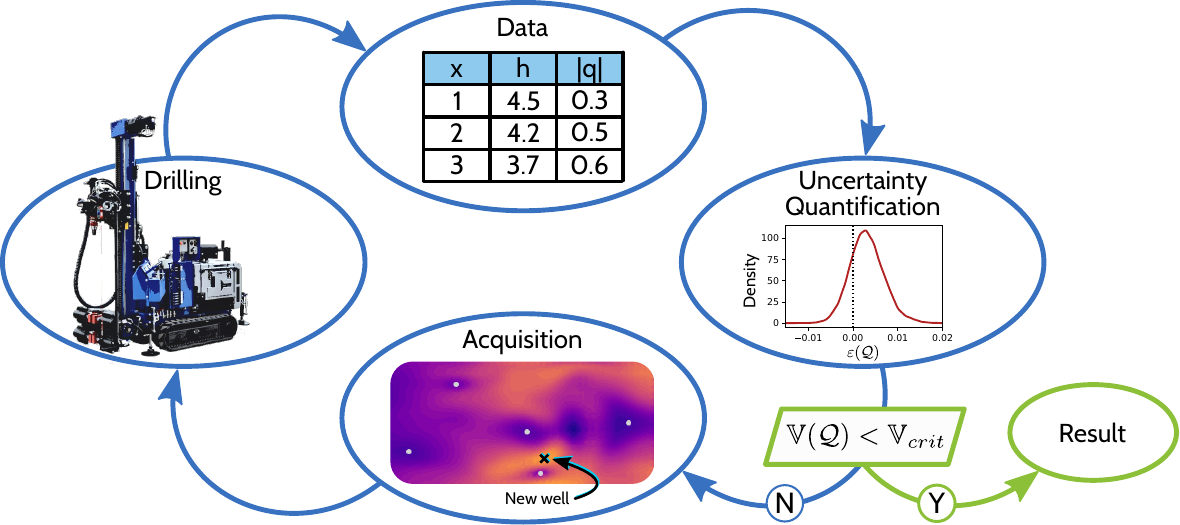}
  \caption{Conceptual diagram of the proposed adaptive optimal design workflow. Here, $\mathbb V(\mathcal Q)$ denotes the variance of the quantity of interest $\mathcal Q$ and $\mathbb V_{crit}$ the desired critical variance.}
  \label{fig:concept}
\end{figure}

Figure \ref{fig:concept} illustrates the proposed workflow at a high level, where new wells are sequentially established at locations of high uncertainty and influence on a QoI, as dictated by the acquisition function. This paper is mainly concerned with the construction of optimal acquisition functions based on the posterior information which would be immediately available from quantifying the uncertainty of the Bayesian inverse problem.

In the following sections, we briefly summarise the theory of Bayesian inverse problems, MCMC and groundwater flow modelling. We then outline the proposed methodology and demonstrate the effectiveness of methodology on a synthetic example. We show that efficient acquisition functions can easily be constructed from information that would already be available from solving the Bayesian inverse problem using MCMC. The method avoids many of the complex calculations that are associated with classic optimal design and exploits information about the Bayesian posterior in a direct and straightforward way.

\section{Theory}
In this section, we first briefly outline the framework of Bayesian inverse problems and Markov Chain Monte Carlo (MCMC), a popular technique employed to draw samples from the Bayesian posterior. We then summarise the fundamentals of groundwater flow modelling for steady-state groundwater flow in a confined aquifer using the Finite Element Method (FEM). Finally, we describe our novel approach to adaptive optimal design of groundwater surveys.

\subsection{Bayesian Inversion} \label{sec:bayes_inv}
A Bayesian inverse problem can be stated compactly as: Given some data $\mathbf{d}$, find the distribution $\pi(\theta \vert \mathbf{d})$ with model parameters $\theta \in \Theta$, where $\Theta$ is the parameter space, so that

\begin{equation} \label{eq:forward}
    \mathbf{d} = \mathcal F(\theta) + \mathbf{\epsilon}
\end{equation}

where $\mathcal F(\theta)$ is the model output and $\mathbf{\epsilon}$ is the measurement error, which is typically assumed to be Gaussian. Bayes theorem then states that 

\begin{equation}
    \pi(\theta \vert \mathbf{d}) = \frac{\pi_{\text{p}}(\theta) \mathcal{L}(\mathbf{d} \vert \theta)}{\pi(\mathbf{d})}
\end{equation}

where $\pi(\theta \vert \mathbf{d})$ is referred to as the \textit{posterior} distribution, $\pi_{\text{p}}(\theta)$ is \textit{prior} distribution, encapsulating what we already know about our model parameters and $\mathcal{L}(\mathbf{d} \vert \theta)$ is called the \textit{likelihood}, essentially a measure of misfit between the model output $\mathcal F(\theta)$ and the data $\mathbf{d}$. While the the so-called \textit{evidence} $\pi(\mathbf{d}) = \int_\Theta \pi_{\text{p}}(\theta) \: \mathcal{L}(\mathbf{d} \vert \theta) \: d\theta$ is generally infeasible or impossible to determine in most real-world scenarios, various sampling techniques allows us to make statistical inferences from $\pi(\theta \vert \mathbf d)$ anyway. Examples include Importance Sampling (IS) and Markov Chain Monte Carlo (MCMC) methods. While these methods are not the object of this study, a short summary of the main ideas of MCMC, which is the specific method employed for inversion in this study, is provided for completeness. 

In MCMC we exploit that $\pi(\mathbf{d})$ is constant and does not depend on the parameters $\theta$. We can therefore write
\begin{equation}
    \pi(\theta \vert \mathbf{d}) \propto \pi_{\text{p}}(\theta) \mathcal{L}(\mathbf{d} \vert \theta)
\end{equation}
or equivalently, for $x, y \in \Theta$
\begin{equation} \label{eq:likelihood_ratio}
    \frac{\pi(y \vert \mathbf{d})}{\pi(x \vert \mathbf{d})} = \frac{\pi_{\text{p}}(y) \mathcal{L}(\mathbf{d} \vert y)}{\pi_{\text{p}}(x) \mathcal{L}(\mathbf{d} \vert x)}
\end{equation}
We then introduce a \textit{transition kernel} or \textit{proposal distribution} $q(y \vert x)$, allowing us to transition from one state $x$ to another $y$. Repeatedly applying the transition kernel $q(y \vert x)$ followed by an accept/reject step prescribed by equation (\ref{eq:acceptance}) we construct a Markov chain where the samples, after an initial \textit{burn-in}, are precisely from the required distribution $\pi(\theta \vert \mathbf{d})$. Here, burn-in refers to the initial MCMC samples which are discarded, since they may not be representative of the equilibrium distribution of the Markov chain. This procedure is described in the box below \citep{metropolis_equation_1953, hastings_monte_1970, gelman_bayesian_2004}.

\begin{center}
    \fbox{\parbox{0.95\textwidth}{
            \textbf{The Metropolis-Hastings Algorithm}, \hspace{0.2cm} $\theta^{(0)} \sim \pi_{\text{p}}(\theta)$, for $i = 0, \dots, N$:
            \begin{enumerate}
                \item Given a parameter realisation $\theta^{(i)}$ and a transition kernel $q(\theta'  \vert  \theta^{(i)})$, generate a proposal $\theta'$.
                \item Compute the acceptance probability of the proposal given the previous realisation: 
                \begin{equation}\label{eq:acceptance}
                \alpha(\theta'|\theta^{(i)}) = \text{min} \left\{ 1, \frac{\pi_{\text{p}}(\theta') \mathcal{L}(\mathbf{d}  \vert  \theta')}{\pi_{\text{p}}(\theta^{(i)}) \mathcal{L}(\mathbf{d}  \vert  \theta^{(i)})} \frac{q(\theta^{(i)} \vert \theta')}{q(\theta' \vert \theta^{(i)})} \right\}
                \end{equation} 
                \item If $u \sim U(0,1) > \alpha$ then set $\theta^{(i+1)} = \theta^{(i)}$, otherwise, set $\theta^{(i+1)} = \theta'$.
            \end{enumerate}
    }}
\end{center}
The acceptance probability (Eq.~\ref{eq:acceptance}) ensures that the algorithm is in detailed balance with the target (posterior) distribution $\pi(\theta \vert \mathbf{d})$. See e.g. \citet[Sec. 5.3]{liu_monte_2004} for more details.
Note that when the measurement error $\mathbf{\epsilon}$ is Gaussian, $\mathbf{\epsilon} \sim \mathcal N(0, \Sigma_\epsilon)$, which we assume in the experiment in Section \ref{sec:example}, then the (unnormalised) likelihood functional takes the following form:

\begin{equation}
    \mathcal{L}(\mathbf{d} \vert \theta) \propto \exp \left( -\frac{1}{2}(\mathcal F(\theta) - \mathbf{d})^T \Sigma_\epsilon^{-1} (\mathcal F(\theta) - \mathbf{d}) \right).
\end{equation}

In this study we employ a number of extensions to the Metropolis-Hastings algorithm to speed up inference, namely the Delayed Acceptance (DA, \citep{christen_markov_2005}) algorithm with finite subchains \citep{lykkegaard_multilevel_2020, lykkegaard2022multilevel}, also referred to as the \textit{surrogate transition method} by \citet{liu_monte_2004}. The DA algorithm exploits an approximate forward model (or Reduced Order Model, ROM) $\hat{\mathcal F}$ to filter MCMC proposals before evaluating them with the fully resolved forward model $\mathcal F$, resulting in a reduction in computational cost. Moreover, we employ a state-independent Approximation Error Model (AEM) to probabilistically correct for model reduction errors introduced by the approximate model, as described by \citet{cui_posteriori_2018}. Finally, we use the Adaptive Metropolis (AM) algorithm as the transition kernel \citep{haario_adaptive_2001}. In this work, we used the open-source DA MCMC framework \texttt{tinyDA}\footnote{\href{https://github.com/mikkelbue/tinyDA}{https://github.com/mikkelbue/tinyDA}} to perform the MCMC sampling.

\subsection{Groundwater Flow} \label{sec:gw_flow}
The groundwater flow equation for steady flow in a confined, inhomogeneous aquifer occupying the domain $\Omega$ with boundary $\Gamma$ can be written as the scalar elliptic partial differential equation
\begin{equation}\label{eq:spde}
    -\nabla \cdot k(\mathbf{x})\nabla u(\mathbf{x}) = g(\mathbf{x}), \quad \mbox{for all} \quad \mathbf{x} \in \Omega,
\end{equation}
subject to boundary conditions on $\Gamma = \Gamma_D \cup \Gamma_N$ with the constraints
\begin{equation}
    u(\mathbf{x}) = u_D(\mathbf{x}) \quad \text{on} \quad \Gamma_D \quad \text{and} \quad -(k(\mathbf{x})\nabla u(\mathbf{x})) \cdot \mathbf{n} = q_N(\mathbf{x}) \quad \text{on} \quad \Gamma_N.
\end{equation}

Here, $k(\mathbf{x})$ is the hydraulic conductivity, $u(\mathbf{x})$ is the hydraulic head, $g(\mathbf{x})$ are sources and sinks, and $\Gamma_D$ and $\Gamma_N$ are boundaries with Dirichlet and Neumann conditions, respectively (see e.g. \citet{diersch_feflow:_2014}). If $\theta$ somehow parameterises the conductivity, then we have $k(\mathbf{x}) = k(\mathbf{x}, \theta)$. This equation can be converted into the weak form by multiplying with a test function $v \in H^1(\Omega)$ and integrating by parts:

\begin{equation} \label{eq:weak_form}
    \int_\Omega \nabla v \cdot (k(\mathbf{x}, \theta) \cdot \nabla u) \: d\mathbf{x} + \int_{\Gamma_N} v \: q_N(\mathbf{x}) \: ds \: = \int_\Omega v \: g({\bf x}) \:  d\mathbf{x}, \quad \forall v\in H^1(\Omega)
\end{equation}
subject to the boundary condition $u(\mathbf{x}) = u_D(\mathbf{x}) \: \text{on} \: \Gamma_D$, where $H^1(\Omega)$ is the Hilbert space of weakly differentiable functions on $\Omega$. We approximate the solution $u(\mathbf{x})$ in a finite element space $V_\tau \subset H^1(\Omega)$ on a finite element mesh $\mathcal Q_\tau(\Omega)$, defined by piecewise linear Lagrange polynomials $\{\phi_i(\mathbf{x})\}_{i=1}^M$ associated with the $M$ finite element nodes. This can be rewritten as a sparse system of equations

\begin{align} \label{eq:gw_se}
    \mathbf{A}(\theta)\mathbf{u} = \mathbf{b} \quad \text{where} \quad 
    A_{ij} &= \int_\Omega \nabla \phi_i(\mathbf{x}) \cdot k(\mathbf{x}, \theta) \nabla \phi_j(\mathbf{x}) d\mathbf{x} \quad \text{and} \\
    b_{i} &= - \int_{\Gamma_N} \phi_i(\mathbf{x})  \: q_N(\mathbf{x}) \: ds \: + \int_\Omega \phi_i(\mathbf{x})  \: g({\bf x}) \:  d\mathbf{x}
\end{align}
where $\mathbf{A}(\theta) \in \mathbb R^{M \times M}$ is the global stiffness matrix and $\mathbf{b} \in \mathbb R^{M}$ is the load vector. The solution to this system $\mathbf{u} := [u_1, u_2, \dots, u_M] \in \mathbb R^M$ represents the hydraulic head at each node, which can be interpolated to the entire domain using the finite element shape functions: $u(\mathbf{x}) = \sum_{i=1}^M u_i \phi_i(\mathbf{x})$. In our numerical experiments, we used the open-source high-performance finite elements package \texttt{FEniCS} \citep{langtangen_solving_2016} to solve these equations. 

\subsection{Adaptive Optimal Design} \label{sec:ada_design}
The overarching research question of this paper is this: if we want to collect more data to reduce the variance in our posterior Monte Carlo estimates, where in the modelling domain $\Omega$ should we do it, to maximise the benefit of the new borehole? More formally, if we let $t$ denote the current design of the survey, so that $\mathbf{d}_t$ and $\pi_t(\theta \vert {\mathbf{d}_t})$ denote, respectively, the data and posterior distribution corresponding to that design, we want to find the next sampling point $\mathbf{x}^\star$ that constrains $\pi_{t+1}(\theta \vert {\mathbf{d}_{t+1}})$ in an optimal way, after setting $\mathbf{d}_{t+1} = (\mathbf{d}_t, d^\star)^T$, where $d^\star$ is the newly collected data at $\mathbf{x}^\star$.

\subsubsection{``Vanilla'' Approach} \label{sec:vanilla}
As outlined in section \ref{sec:bayes_inv}, Bayesian inversion allows us to construct the posterior distribution of parameters given the data $\pi_t(\theta \vert \mathbf{d}_t)$. If the inversion was completed using MCMC, and obtaining the model output $\mathcal F(\theta)$ involved solving some partial differential equation with solution $u(\mathbf{x})$, we can cache these solutions during sampling, and would after sampling possess a set of pairs $\{(\theta^{(i)}, u^{(i)}(\mathbf{x}))\}_{i=0}^{N^\dagger}$. Since $\{\theta^{(i)}\}_{i=0}^{N^\dagger}$ are distributed exactly according to $\pi_t(\theta \vert \mathbf{d}_t)$, so are any functions of $\theta$, such as $u(\mathbf{x})$. Here, $N^\dagger$ is the number of MCMC samples after discarding the burn-in. Hence, we can easily obtain Monte Carlo estimates for

\begin{equation*}
    \mathbb E_{\pi_t(\theta \vert \mathbf{d}_t)}[u(\mathbf{x},\theta)] \quad \mbox{and} \quad \mathbb D_{\pi_t(\theta \vert \mathbf{d}_t)}[u(\mathbf{x},\theta)]
\end{equation*}

Here, $\mathbb D$ signifies some measure of statistical dispersion, for example variance, standard deviation, or entropy. We could, in accordance with the maximum entropy approach \citep{shewry_maximum_1987}, postulate that the accuracy of our inversion is driven by the dispersion in $u({\bf x})$ and hence we could solve the following optimisation problem
\begin{equation} \label{eq:vanilla}
    \mathbf{x}^\star = \argmax_{\mathbf{x} \in \Omega} \mathbb D_{\pi_t(\theta \vert \mathbf{d}_t)}[u(\mathbf{x},\theta)]
\end{equation}

\subsubsection{Dual-Weighted Approach} \label{sec:dual_weighted}
The simple approach outlined above will improve the general quality of $u(\mathbf{x})$, but it is limited by the fact that it is not tailored for a particular quantity of interest $\mathcal Q$ and this is where the \textit{dual weighted} approach comes into play. In this context, rather than simply sampling from places with high uncertainty, we aim to pick sampling points that also have a high expected influence on our quantity of interest $\mathcal Q$. This is exactly the problem, that \textit{adjoint} or \textit{dual} state methods aim to solve \citep{plessix_review_2006}.

Suppose in a particular application, we are interested in estimating a particular quantity of interest $\mathcal Q(u)$, which we can write as a functional of the solution. For example, if our quantity of interest is the hydraulic head around a point $\mathbf{x}' \in \Omega$, we could choose
\begin{equation}
    \mathcal Q_{\mathbf{x}'}(u) = \int_\Omega u(\mathbf{x}) \exp\left(-\frac{(\mathbf{x} - \mathbf{x}')^2}{\lambda}\right)\;d\mathbf{x}
\end{equation}
for some sufficiently small length scale $\lambda$. This, however, is a trivial problem, since if the quantity of interest is the hydraulic head at some point, we can just place our monitoring well at that point and measure it. It would be much more useful to target a quantity of interest that we cannot measure directly. Hence, in this study we consider flux over a boundary $\Gamma'$ with the following functional:
\begin{equation} \label{eq:q_boundary}
    \mathcal Q_{\Gamma'}(u) = \int_{\Gamma'} [-k(\mathbf{x}, \theta) \cdot \nabla u(\mathbf{x})] \cdot \mathbf{n} \: ds
\end{equation}

\noindent The adjoint state equation associated with Eq.~(\ref{eq:q_boundary}) is
\begin{equation} \label{eq:adjoint_state_equation}
    \nabla \cdot k \nabla \omega = 0
\end{equation}
subject to the boundary conditions 
\begin{align*}
\omega_{D}(\mathbf{x})       &= 0 &\text{on} \quad &\Gamma_D\setminus\Gamma'  \\
\omega_{\Gamma'}(\mathbf{x}) &= 1 &\text{on} \quad &\Gamma'\\
q_N^\omega(\mathbf{x})       &= (k(\mathbf{x})\nabla \omega(\mathbf{x})) \cdot \mathbf{n} = 0 &\text{on} \quad &\Gamma_N.
\end{align*}
The solution $\omega(\mathbf{x})$ is called the adjoint state or \textit{influence} function. Please refer to \citet{sykes_sensitivity_1985} and \ref{ap:adjoint_equation} for details on the derivation of the adjoint state equation and its associated boundary conditions. Integrating by parts and multiplying with a test function $v \in H^1(\Omega)$, we arrive at the weak form of the adjoint state equation:
\begin{equation} \label{eq:adjoint}
    \int_\Omega \nabla v \cdot (k(\mathbf{x}, \theta) \cdot \nabla \omega) \: d\mathbf{x} + \int_{\Gamma_N} v \: q_N^\omega(\mathbf{x}) \: ds \: = 0, \; \forall v\in H^1(\Omega)
\end{equation}
subject to boundary conditions $\omega_{D}(\mathbf{x}) = 0 \: \text{on} \: \Gamma_D\setminus\Gamma' \: \text{and} \: \omega_{\Gamma'}(\mathbf{x}) = 1 \: \text{on} \: \Gamma'$. Given some conductivity parameters $\theta$, \eqref{eq:adjoint} can be discretised using the same finite element grid as \eqref{eq:gw_se}, leading to the following sparse system of equations:

\begin{align}
    \mathbf{A}(\theta)\mathbf{\omega} = \mathbf{b}_{\omega} \quad \text{where} \quad 
    A_{ij} &= \int_\Omega \nabla \phi_i(\mathbf{x}) \cdot k(\mathbf{x}, \theta) \nabla \phi_j(\mathbf{x}) d\mathbf{x} \quad \text{and} \\
    b_{\omega, i} &= - \int_{\Gamma_N} \phi_i(\mathbf{x})  \: q_N^\omega(\mathbf{x}) \: ds.
\end{align}
It is important to note here, that the stiffness matrix $\mathbf{A}(\theta)$, since the steady-state groundwater flow equation is \textit{self-adjoint}, is exactly the same as in equation (\ref{eq:gw_se}), and the assembled system can hence be partially recycled when solving both equations. However, since the boundary conditions for the adjoint state equation are different than for the primal problem, care must be taken when assembling the adjoint system of equations. After solving this system of equations, the influence function can be interpolated to the entire domain using our finite element shape functions:

\begin{equation*}
    \omega(\mathbf{x}) = \sum_{i=1}^M\omega_i\phi_i(\mathbf{x}) \quad \mbox{where} \quad \mathbf{\omega} = [\omega_1, \omega_2, \ldots, \omega_M]^T.
\end{equation*}
The influence function is commonly interpreted as the sensitivity of the quantity of interest to a unit point source anywhere on the domain \citep{sykes_sensitivity_1985, wilson_illustration_1985}, or in this particular case as the sensitivity of flow anywhere on the domain to the boundary condition. Broadly speaking, the influence function directs us towards areas of the modelling domain with a potentially high influence on our quantity of interest, which is what we required for our dual-weighted approach.

\begin{figure}[htbp]
  \centering
  \includegraphics[width=0.5\textwidth]{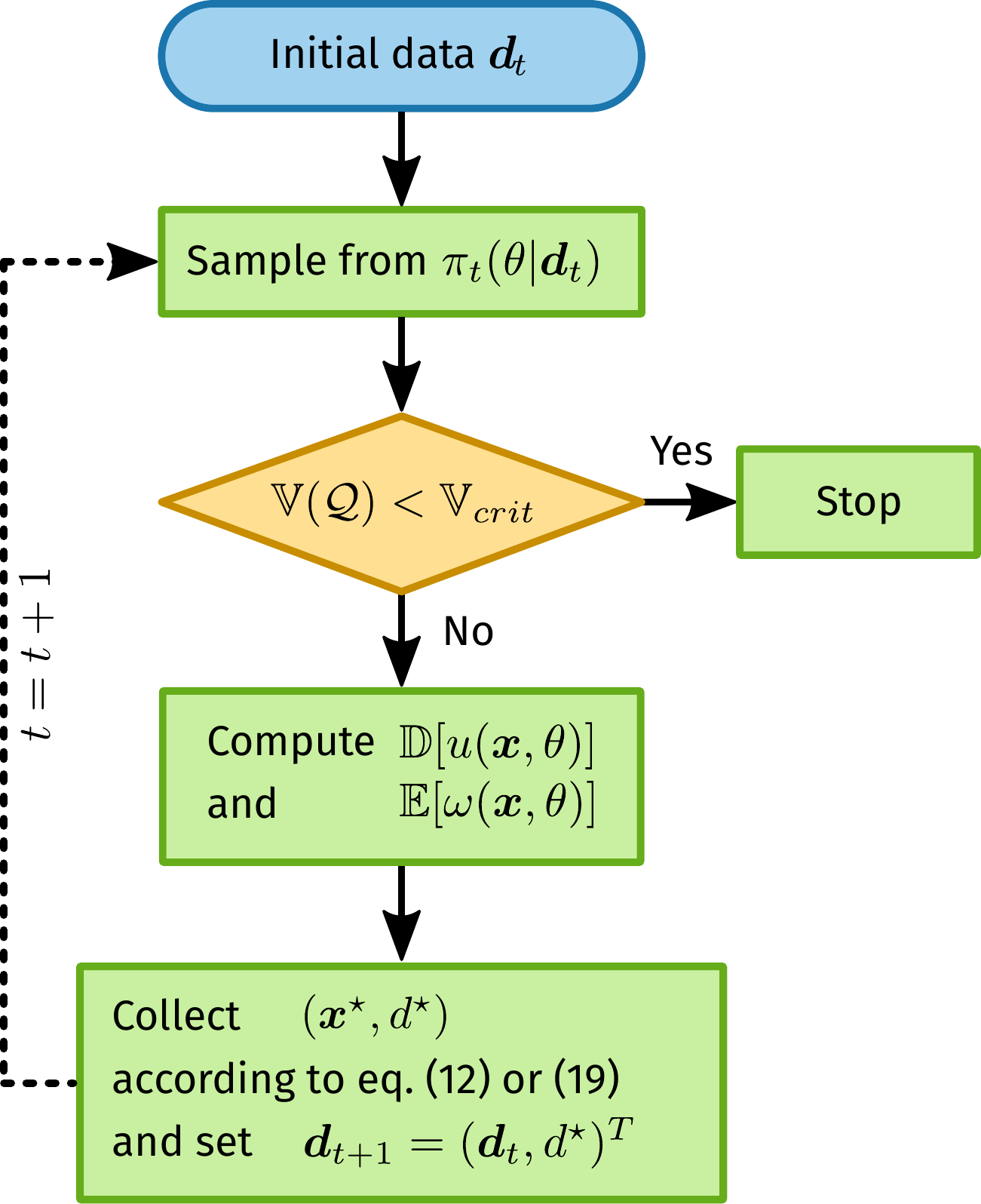}
  \caption{Proposed adaptive optimal design procedure. As in Figure \ref{fig:concept}, $\mathbb V(\mathcal Q)$ denotes the variance of the quantity of interest $\mathcal Q$ and $\mathbb V_{crit}$ the desired critical variance.}
  \label{fig:procedure}
\end{figure}

We note that $\omega(\mathbf{x})$ is now a random function which depends on model parameters $\theta$, and we can obtain estimates for $\mathbb E_{\pi_t(\theta \vert \mathbf{d}_t)}[\omega({\bf x},\theta)]$. Hence, we propose the following acquisition function

\begin{equation} \label{eq:dual_weighted}
    \mathbf{x}^\star = \argmax_{\mathbf{x} \in \Omega} \: \mathbb D_{\pi_t(\theta \vert \mathbf{d}_t)}[u(\mathbf{x},\theta)] \cdot |  \mathbb E_{\pi_t(\theta \vert \mathbf{d}_t)}[\omega(\mathbf{x},\theta)] |.
\end{equation}

where $|\cdot|$ denotes the absolute value. We use the absolute value of the expectation of the influence function to make sure that the weighting is always positive, since $\omega(\mathbf{x, \theta})$ is not always positive for other adjoint equations. We call this approach dual-weighted, since we are essential re-weighting the dispersion $\mathbb D_{\pi_t(\theta \vert \mathbf{d}_t)}[u(\mathbf{x},\theta)]$, by the expected solution of the dual problem. Figure \ref{fig:procedure} illustrates the different steps in the proposed adaptive optimal design procedure.

\subsubsection{Remarks} \label{sec:remarks}
(1) The dual-weighted approach can be considered a hybrid between the goal-oriented error estimation employed for mesh-adaptation in the context of various expensive and mesh-sensitive PDE problems (see e.g. \citet{prudhomme_goal-oriented_1999, oden_goal-oriented_2001}), and Bayesian Optimisation (BO), typically used to optimise some unknown function approximated with sparse and/or noisy data (see e.g. \citet{mockus_bayesian_1989, frazier_tutorial_2018}). In this context, our dual-weighted approach could be framed as a form of prior-guided BO \citep{souza_bayesian_2021}, where $\omega(\mathbf{x})$ broadly represents our prior belief that any point $\mathbf{x}$ constitutes a ``good'' sampling location. However, we remark that in our formulation $\omega(\mathbf{x})$ is not a probability distribution but a random weighting function.

(2) In the above formulations, we have chosen the dispersion of the hydraulic head $\mathbb D_{\pi_t(\theta \vert \mathbf{d}_t)}[u(\mathbf{x},\theta)]$ as the function representing uncertainty in the model. Other sensible choices of uncertainty metrics would be the dispersion of the hydraulic conductivity $\mathbb D_{\pi_t(\theta \vert \mathbf{d}_t)}[k(\mathbf{x},\theta)]$, or of some norm of the flux $\mathbb D_{\pi_t(\theta \vert \mathbf{d}_t)}[\lVert \mathbf{q}(\mathbf{x},\theta) \rVert_p]$.

(3) Since sampling from $\pi_t(\theta \vert \mathbf{d}_t)$ can be computationally expensive, it may be desirable to pick multiple new sampling locations at each step of the algorithm. Denote the number of new sampling locations in each such batch acquisition as $N^\star$. Then this can be achieved by penalising the acquisition function by some local penalisation functions $\{\psi_{\mathbf{x}^\star_{i}}(\mathbf{x})\}_{i=1}^{N^\star-1}$, centered on the previous sampling points $\{\mathbf{x}^\star_{i}\}_{i=1}^{N^\star-1}$ of the current batch, as described in \citet{pmlr-v51-gonzalez16a}. This approach would yield the following dual-weighted batch acquisition function for $\{\mathbf{x}^\star_{i}\}_{i=2}^{N}$:

\begin{equation} \label{eq:dual_weighted_batch}
    \mathbf{x}_{i}^\star = \argmax_{\mathbf{x} \in \Omega} \: \mathbb D_{\pi_t(\theta \vert \mathbf{d}_t)}[u(\mathbf{x},\theta)] \cdot | \mathbb E_{\pi_t(\theta \vert \mathbf{d}_t)}[\omega(\mathbf{x},\theta)] | \cdot \prod_{j=1}^{i-1} \psi_{\mathbf{x}^\star_{j}}(\mathbf{x}).
\end{equation}
Similarly, the batch acquisition function for the vanilla approach takes the form
\begin{equation} \label{eq:vanilla_batch}
    \mathbf{x}_{i}^\star = \argmax_{\mathbf{x} \in \Omega} \: \mathbb D_{\pi_t(\theta \vert \mathbf{d}_t)}[u(\mathbf{x},\theta)] \cdot \prod_{j=1}^{i-1} \psi_{\mathbf{x}^\star_{j}}(\mathbf{x}).
\end{equation}
A reasonable choice of penalisation functions would be the Gaussian \begin{equation}\psi_{\mathbf{x}'}(\mathbf{x}) = 1 - \exp\left(-\frac{1}{2}\frac{\lVert \mathbf{x} - \mathbf{x}'\lVert_2^2}{l_\psi} \right)
\end{equation}
where $l_\psi$ controls the dispersion of the function and $\lVert \cdot \lVert_2$ is the $L^2$-norm. Using such a penalisation function, the acquisition function would be exactly zero at previous sampling points from the current batch, and smoothly rebound to Eq.~(\ref{eq:dual_weighted}) or Eq.~(\ref{eq:vanilla}) as the distance to previous sampling points increases.

(4) As mentioned earlier, we formulate our method in the context of steady state groundwater flow in a confined aquifer. While this is the most common approach to groundwater flow modelling, it is, naturally, not exhaustive. For a detailed analysis of the adjoint state equations for transient groundwater flow, we refer the to e.g. \citet{sun_inverse_1999} and \citet{lu_analytical_2015}. The unconfined case is considerably more complex, since the constitutive equations are nonlinear. While unconfined groundwater flow can, under some assumptions, be reasonably approximated by the constitutive equations for confined flow \citep{wang_introduction_1982}, this is not always the case. For a derivation and analysis of the adjoint equations pertaining to unconfined and coupled aquifers, we refer to e.g. \citet{sun_inverse_1999} and \citet{neupauer_adjoint_2012}.

(5) Note that the constitutive and adjoint equations are discretised using FEM in the above section. We restrict ourselves to this method for brevity, but remark that the proposed acquisition functions (Eqs.~(\ref{eq:vanilla}), (\ref{eq:dual_weighted}), (\ref{eq:dual_weighted_batch}) and (\ref{eq:vanilla_batch})) are valid for any discretisation scheme. Also note that if piecewise linear shape functions are employed to approximate $u(\mathbf{x})$, the maxima of the acquisition functions will occur at finite element nodes.

\section{Example} \label{sec:example}
In this section, we demonstrate the vanilla and dual-weighted approach in the context of a synthetic groundwater flow example. We first outline the model setup, including the geological model and finite element representation. We then explain the particular methodology for this example in detail. Finally, we present the results.

\subsection{Model Setup}
We model the hydraulic conductivity as a log-Gaussian Random Field with a Matern 3/2 covariance kernel:
\begin{equation} \label{eq:matern}
    C(\mathbf{x}, \mathbf{y}) = \left( 1 + \sqrt{3} \frac{\lVert \mathbf{x} - \mathbf{y} \lVert_2}{l} \right) \exp \left(- \sqrt{3} \frac{\lVert \mathbf{x} - \mathbf{y} \lVert_2}{l} \right)
\end{equation}
where $l$ is the length scale \citep{rasmussen_gaussian_2006} and $\lVert \cdot \lVert_2$ is the $L^2$-norm. The resulting random field is expanded in an orthogonal eigenbasis with $N_{\text{KL}}$ Karhunen–Loève (KL) eigenmodes. To this end, we construct a matrix of covariances between each pair of finite element nodes $\mathbf{C} \in \mathbb R^{M \times M}$ according to Eq.~(\ref{eq:matern}), so that $C_{ij} = C(\mathbf{x}_i, \mathbf{x}_j)$. This covariance matrix $\mathbf{C}$ is decomposed into the $N_{\text{KL}}$ largest eigenvalues $\{\lambda_i\}_{i=1}^{N_{\text{KL}}}$ and eigenvectors $\{\mathbf{\psi}_i\}_{i=1}^{N_{\text{KL}}}$.
The nodal conductivities $\mathbf{k} := [k_1, k_2, \dots, k_M]$ are then given by
\begin{equation}
    \log \mathbf{k} = \mathbf{\mu} + \sigma \mathbf{\Psi} \mathbf{\Lambda}^{\frac{1}{2}} \mathbf{\theta}
\end{equation}
with $\mathbf{\Lambda} = \text{diag}([\lambda_1, \lambda_2, \dots, \lambda_{N_{\text{KL}}}])$ and $\mathbf{\Psi} = [\mathbf{\psi}_1, \mathbf{\psi}_2, \dots, \mathbf{\psi}_{N_{\text{KL}}}]$. The vector $\mathbf{\mu} = \mu\mathbf{1}$ is the mean of the log-conductivity, $\sigma$ is the standard deviation of the log-conductivity, and $\mathbf{\theta} \sim \mathcal N(0, \mathbb I_{N_{\text{KL}}})$ \citep{dodwell_hierarchical_2015}. When defined in this way, the associated Bayesian inverse problem involves exploring  $\pi(\theta \vert \mathbf{d})$, i.e. the posterior distribution of hydraulic conductivity parameters $\theta$ given measurements $\mathbf{d}$, where the aforementioned normal distribution constitutes the prior distribution of parameters: $\pi_{\text{p}}(\mathbf{\theta}) = \mathcal N(0, \mathbb I_{N_{\text{KL}}})$.

\begin{figure}[htbp]
    \begin{subfigure}{0.5\textwidth}
        \centering
        \includegraphics[width=0.95\textwidth]{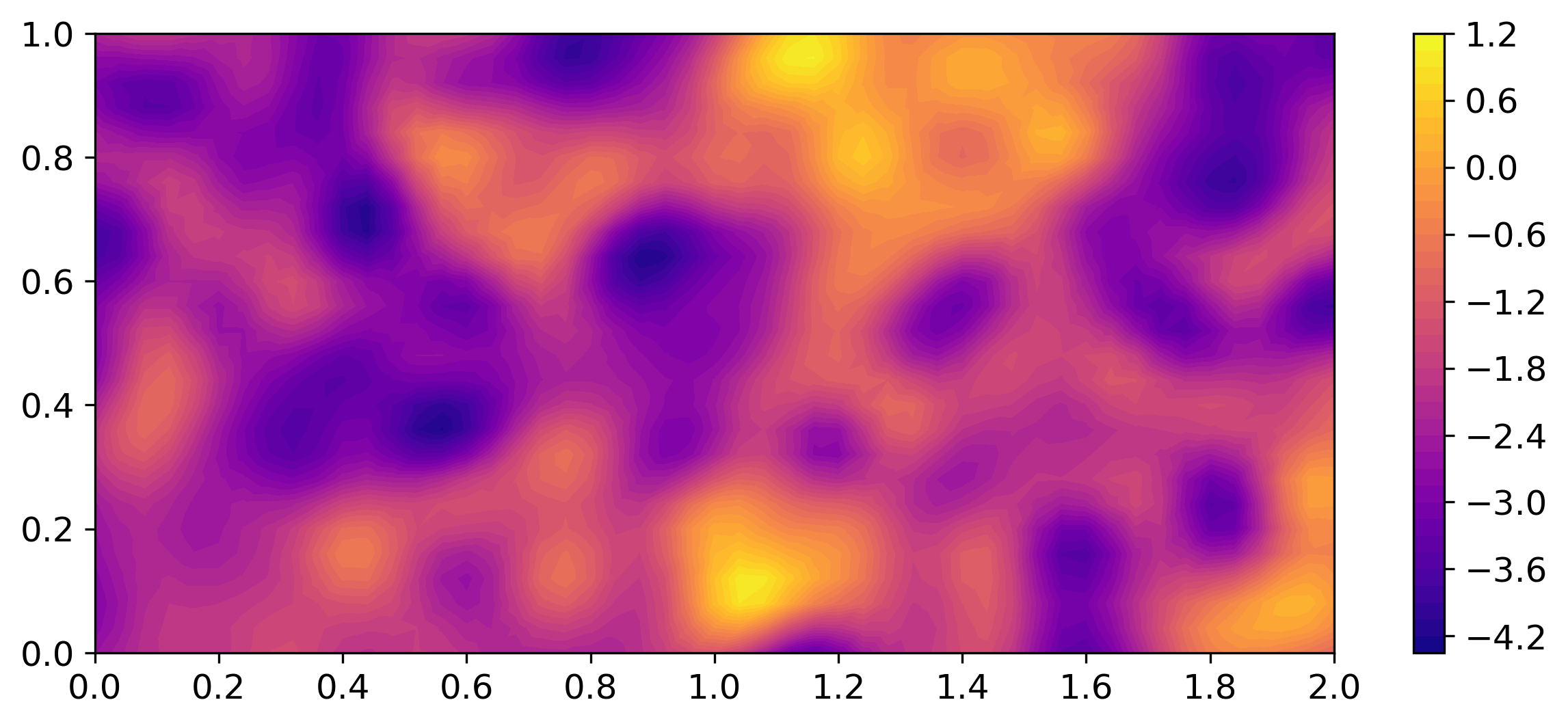}
        \caption{Conductivity for the data-generating model.}
        \label{fig:conductivity_true}
    \end{subfigure}
    \begin{subfigure}{0.5\textwidth}
        \centering
        \includegraphics[width=0.95\textwidth]{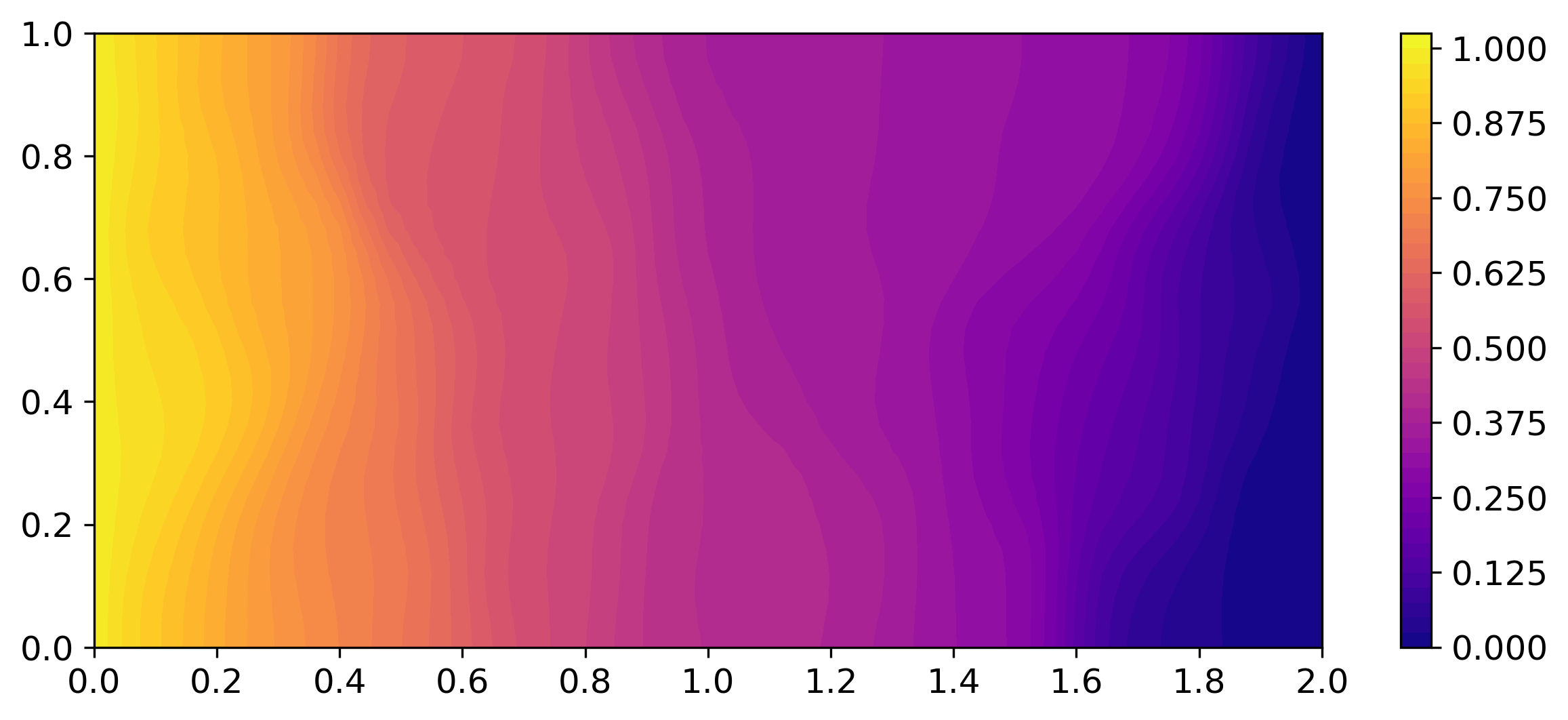}
        \caption{Hydraulic head.}
        \label{fig:head_true}
    \end{subfigure}
    \begin{subfigure}{0.5\textwidth}
        \centering
        \includegraphics[width=0.95\textwidth]{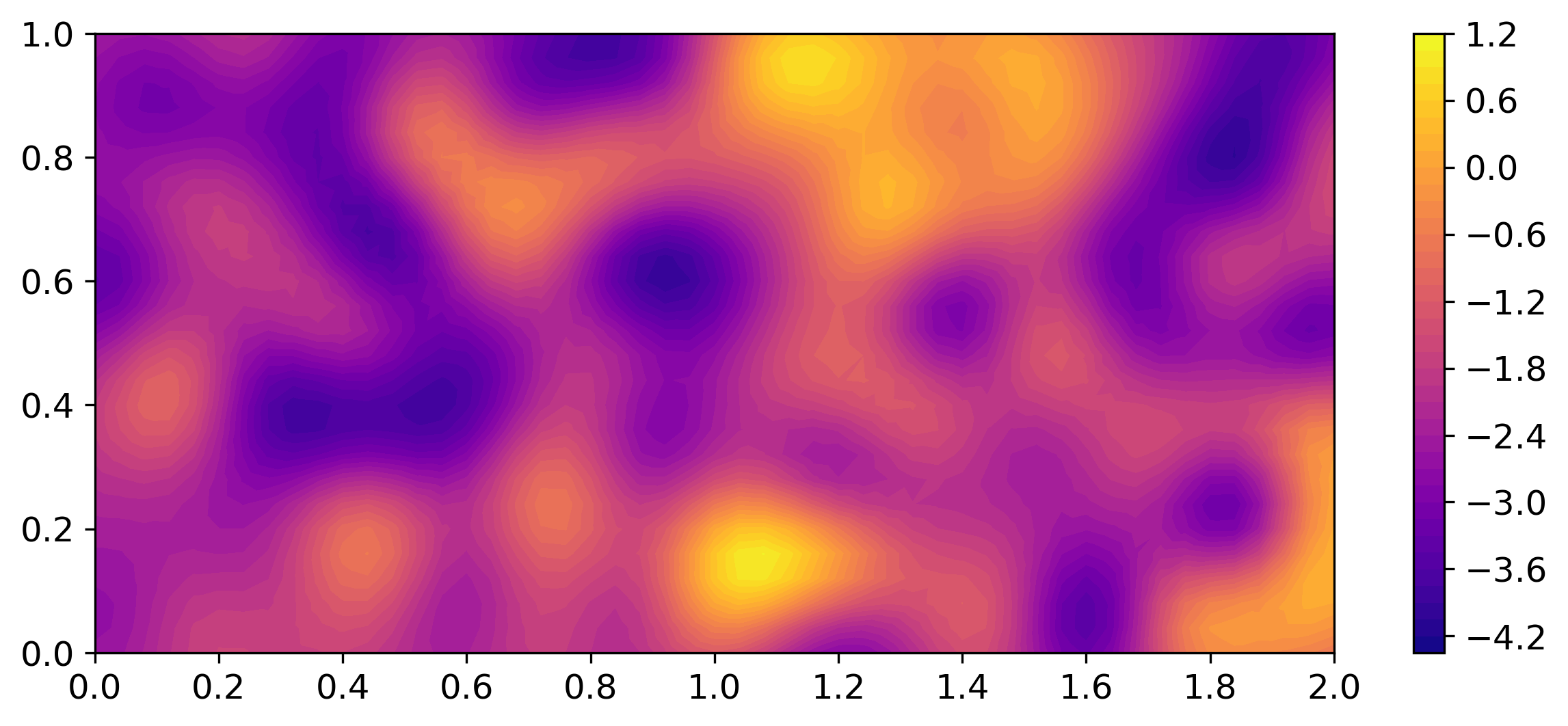}
        \caption{Conductivity for the fine forward model $\mathcal F$.}
        \label{fig:conductivity_fine}
    \end{subfigure}
    \begin{subfigure}{0.5\textwidth}
        \centering
        \includegraphics[width=0.95\textwidth]{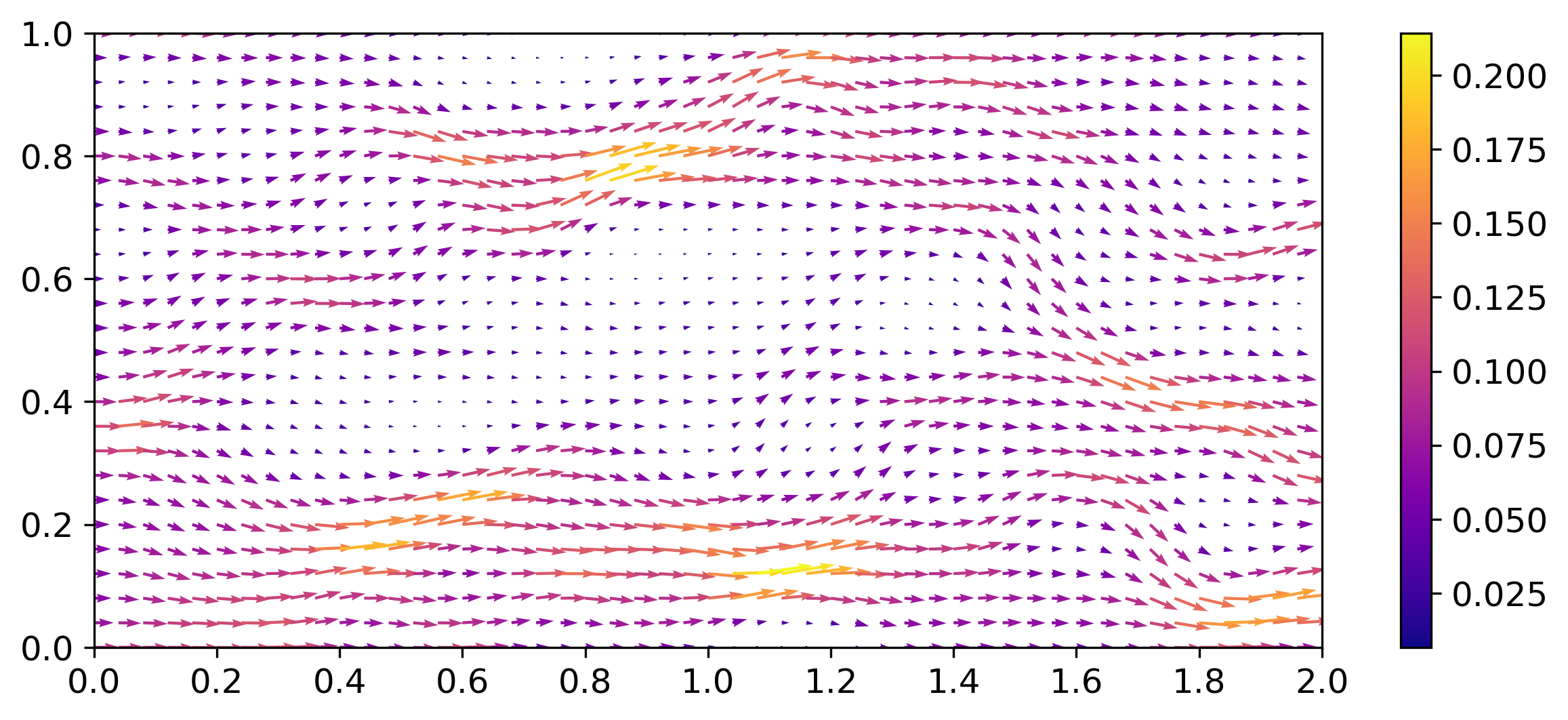}
        \caption{Flux.}
        \label{fig:flux_true}
    \end{subfigure}
    \begin{subfigure}{0.5\textwidth}
        \centering
        \includegraphics[width=0.95\textwidth]{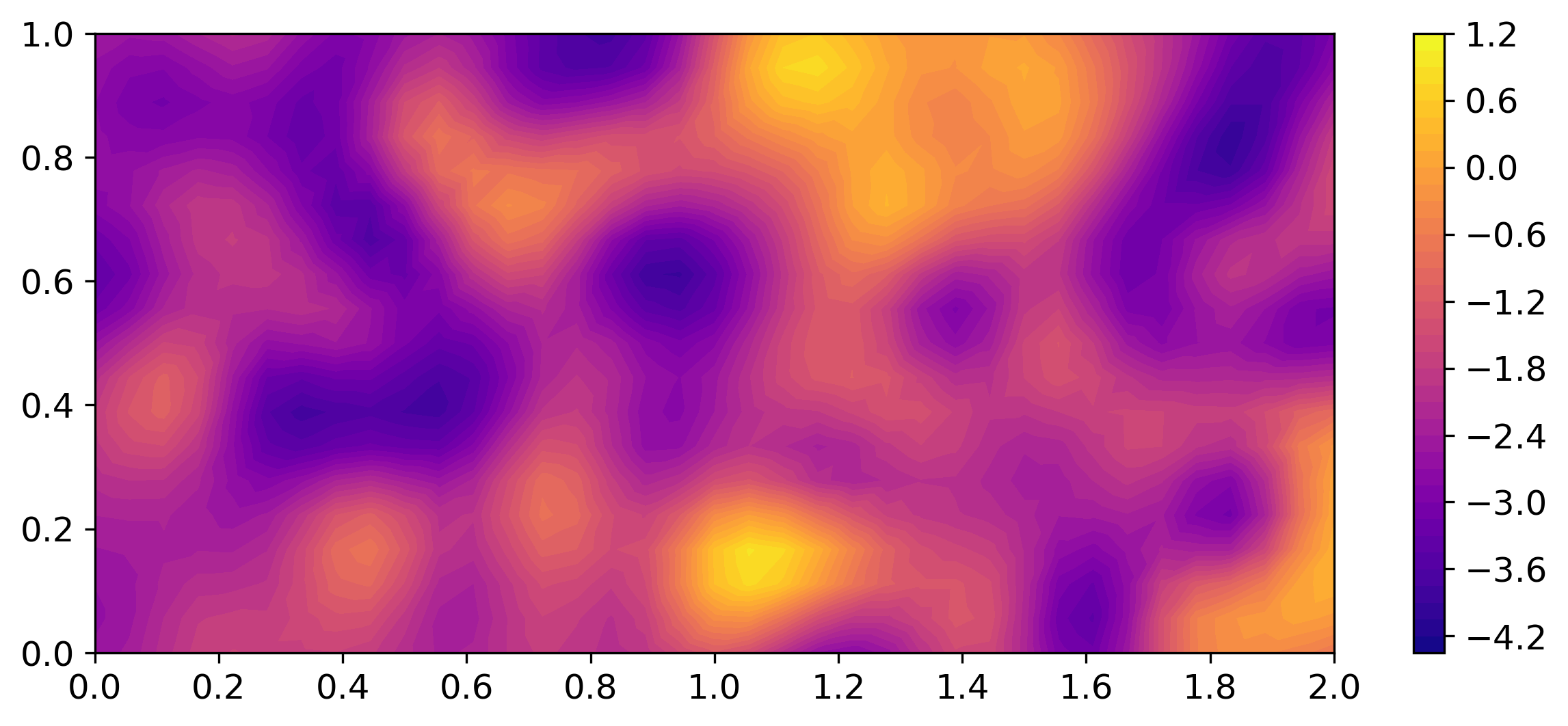}
        \caption{Conductivity for the coarse forward model $\hat{\mathcal F}$.}
        \label{fig:conductivity_coarse}
    \end{subfigure}
    \begin{subfigure}{0.5\textwidth}
        \centering
        \includegraphics[width=0.95\textwidth]{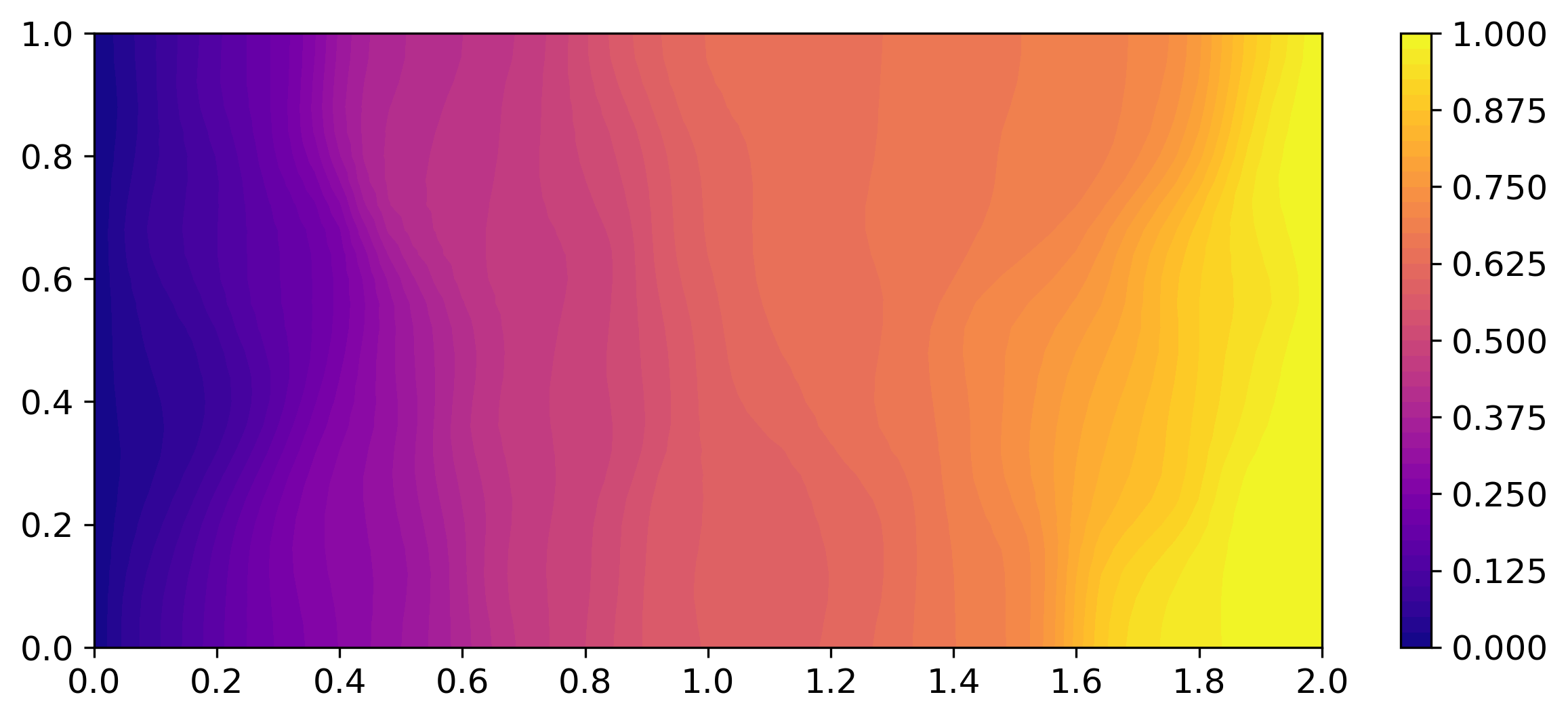}
        \caption{Influence function.}
        \label{fig:influence_true}
    \end{subfigure}%
  \caption{A random realisation from the prior $\pi_{\text{p}}(\mathbf{\theta})$, with the corresponding primary and adjoint solutions. The left column shows the conductivity for the data-generating model (\subref{fig:conductivity_true}), the fine forward model (\subref{fig:conductivity_fine}) and the coarse forward model (\subref{fig:conductivity_coarse}) respectively. The right column shows the hydraulic head (\subref{fig:head_true}), the flux (\subref{fig:flux_true}), and the influence function (\subref{fig:influence_true}), respectively.}
  \label{fig:gaussian_truth}
\end{figure}

We used three different models for the experiments (Fig. \ref{fig:gaussian_truth}), one \textit{data-generating} model representing the ground truth, a \textit{fine} forward model representing the fully resolved forward model $\mathcal F$ in the Bayesian inverse problem (see Eq.~(\ref{eq:forward})), and a \textit{coarse} forward model, corresponding to the reduced order forward model in the Delayed Acceptance MCMC sampler $\hat{\mathcal F}$, as described in e.g. \citet{christen_markov_2005, liu_monte_2004, cui_posteriori_2018, lykkegaard_multilevel_2020, lykkegaard2022multilevel}. Note that using the dual-weighted approach described herein does not require a Delayed Acceptance MCMC sampler. Any method capable of producing Monte Carlo samples from the posterior will do.

The experiments were performed on a rectangular domain $\Omega = [0,2] \times [0,1]$ meshed using a structured triangular grid with $M_{fine} = 1326$ degrees of freedom for the data-generating model and the fine forward model, and $M_{coarse} = 703$ degrees of freedom for the coarse forward model. For the data-generating model, the log-Gaussian random conductivity was truncated at $N_{\text{KL}} = 256$ KL eigenmodes, while for the fine and coarse models it was truncated at $N_{\text{KL}} = 128$. Hence the dimensionality of the inverse problem in these experiments was $128$, which is very high and a challenging problem for any MCMC algorithm. Moreover, we set $l = 0.1$, $\mu = -2$ and $\sigma = 1.0$ for every model. This resulted in strongly anisotropic conductivity fields with log-conductivities broadly between -5 and 1 (Fig. \ref{fig:conductivity_true}).

We imposed fixed head Dirichlet boundary conditions of 1 and 0 on the left and right boundaries, respectively, and no-flow Neumann conditions on the remaining top and bottom boundaries. We set the right hand side of Eq.~(\ref{eq:spde}) to $g(\mathbf{x}) = 0$. We chose flux across the right boundary $\Gamma_r$ as our quantity of interest $\mathcal Q$, corresponding to the following functional (as in equation (\ref{eq:q_boundary})):
\begin{equation}
    \mathcal Q(u) = \int_{\Gamma_r} [-k(\mathbf{x}, \theta) \cdot \nabla u(\mathbf{x})] \cdot \mathbf{n} \: ds
\end{equation}
and the associated adjoint state equation shown in (\ref{eq:adjoint_state_equation}) with $\Gamma' = \Gamma_r$. Figure \ref{fig:influence_true} shows an example of the influence function generated by this adjoint state equation.
The left column of Fig. \ref{fig:gaussian_truth} shows the conductivity associated with a random draw from the prior $\pi_{\text{p}}(\mathbf{\theta})$, for the data-generating model, the fine model, and the coarse model, respectively. The right column of Fig. \ref{fig:gaussian_truth} shows the corresponding hydraulic head, flux and influence function for the data-generating model.

\subsubsection{Methodology}
\begin{figure}[htbp]
    \begin{subfigure}{0.5\textwidth}
        \centering
        \includegraphics[width=0.95\textwidth]{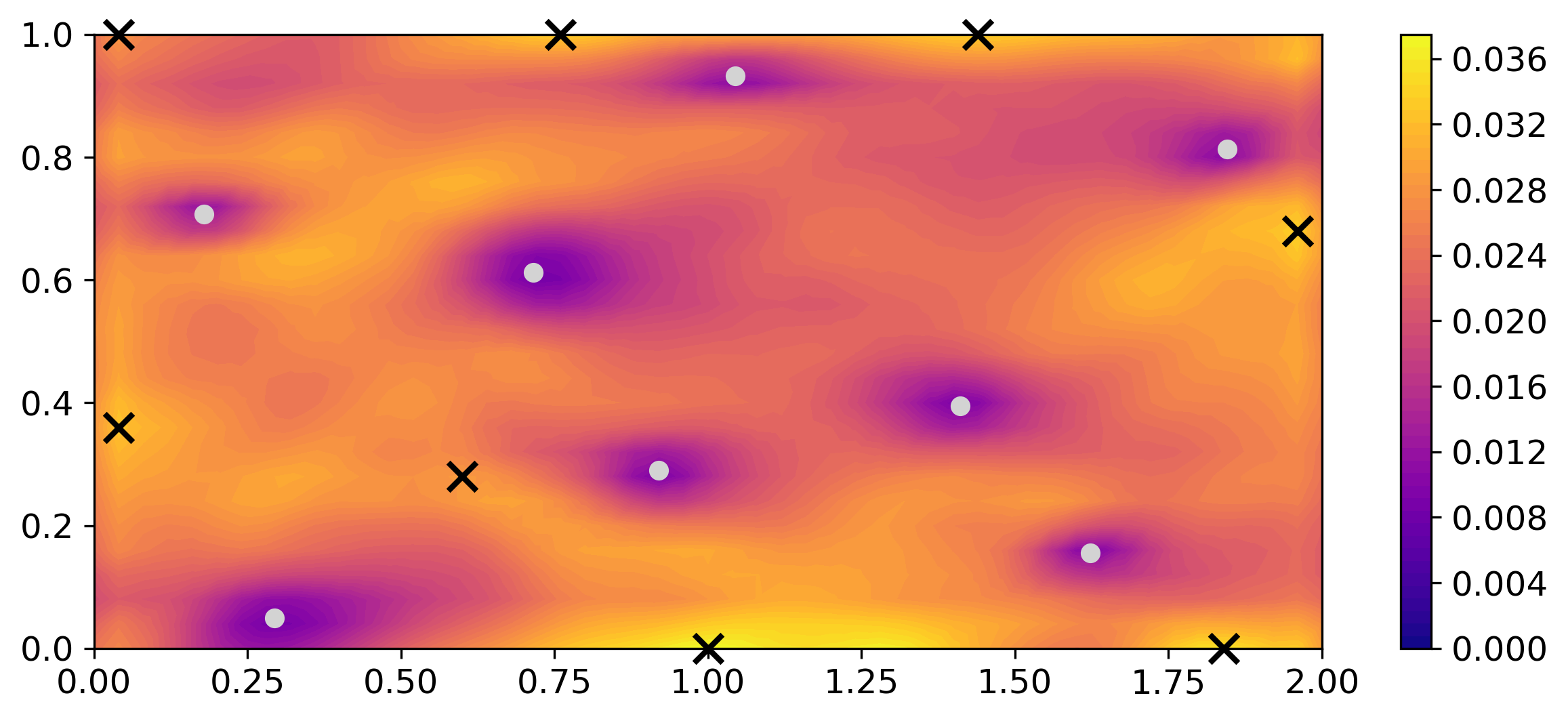}
        \caption{Vanilla acquisition $S(\lVert \mathbf{q}(\mathbf{x}) \rVert_2)$.}
        \label{fig:vanilla}
    \end{subfigure}
    \begin{subfigure}{0.5\textwidth}
        \centering
        \includegraphics[width=0.95\textwidth]{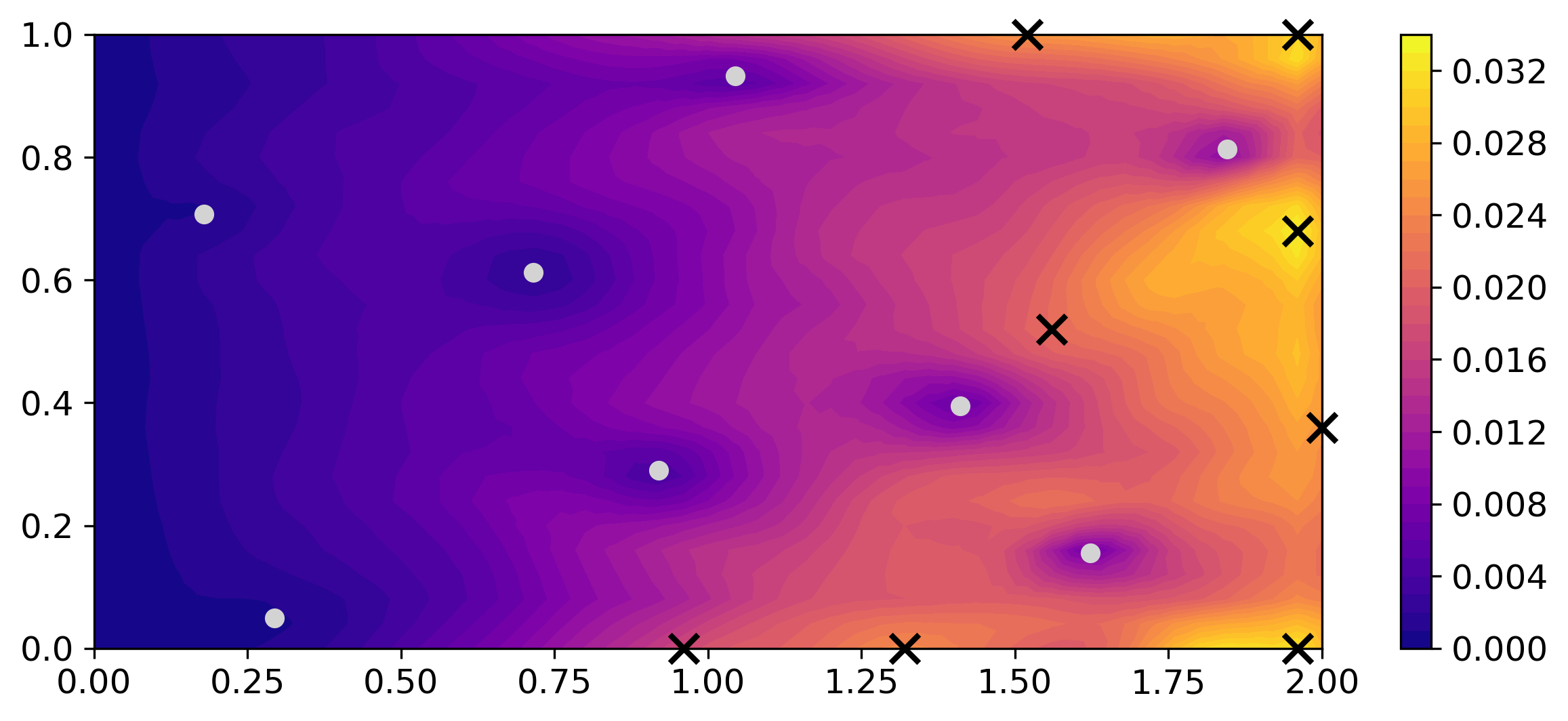}
        \caption{Dual-weighted acquisition $S(\lVert \mathbf{q}(\mathbf{x}) \rVert_2) \cdot \bar{\omega}(\mathbf{x})$.}
        \label{fig:adjoint}
    \end{subfigure}
  \caption{Acquisition functions of the vanilla and dual-weighted approaches for one sample of the $n=30$ models. The white dots show the initial datapoints, while the black crosses show the new datapoints suggested by each acquisition function.}
  \label{fig:acquisition}
\end{figure}

Using the above setup, we completed a total of $n=30$ independent numerical experiments to demonstrate the feasibility of the dual-weighted approach. We chose the standard deviation of the $L^2$-norm of the flux $S(\lVert \mathbf{q}(\mathbf{x}) \rVert_2)$ as the general measure of uncertainty in the model. For each independent experiment, the following experimental procedure was observed: \begin{enumerate*}[label=(\arabic*)] \item The hydraulic conductivity for the data-generating model was initialised with a random draw from the prior, and the primary problem was solved. \item Eight observation wells were placed randomly on the domain by Latin Hypercube sampling \citep{lhs} (see Fig. \ref{fig:acquisition}). \item \label{step:observations} For each observation well $\mathbf{x}_i$, the hydraulic head $u(\mathbf{x}_i)$ and the norm of the flux $\lVert \mathbf{q}(\mathbf{x}_i) \lVert_2$ were computed. These head and flux observations were contaminated with white noise from $\epsilon_u \sim \mathcal N(0, 0.01^2)$ and $\epsilon_{\lVert q \lVert_2} \sim \mathcal N(0, 0.001^2)$, respectively. \item Delayed Acceptance MCMC sampling was completed with $2$ independent samplers each drawing $N=25000$ fine samples with a subsampling length of 5 (see e.g. \citet{lykkegaard_multilevel_2020, lykkegaard2022multilevel}), and a burn-in of $N_{burn} = 5000$ was discarded. This resulted in a total number of MCMC samples of $N^\dagger= 40000$ for each experiment. \item The standard deviation of the $L^2$-norm of the flux $S(\lVert \mathbf{q}(\mathbf{x}) \rVert_2)$ and the mean of the influence function $\bar{\omega}(\mathbf{x})$ were computed at the finite element nodes and interpolated to the entire domain using the finite element shape functions, and eight new observation wells were placed according to the batch vanilla and dual-weighted acquisition functions, see Eq.~(\ref{eq:vanilla_batch}) and Eq.~(\ref{eq:dual_weighted_batch}). Figure \ref{fig:acquisition} shows the vanilla and dual weighted acquisition functions for one sample of the $n=30$ models. As expected, the weighting function $\bar{\omega}(\mathbf{x})$ prioritised observation wells closer to the boundary of the quantity of interest. \item Data were extracted from the four new observation wells as in step \ref{step:observations} and appended to the data vector. \item Delayed Acceptance MCMC sampling was repeated, using the new data vectors for both the vanilla and dual-weighted approaches. \end{enumerate*}

For each experiment and each posterior distribution (initial, vanilla, and dual-weighted) with each $N^\dagger = 40000$ posterior samples, we computed the mean squared error (MSE) and variance of the predicted quantity of interest $\{\mathcal Q^{(i)}\}_{i=1}^{N^\dagger}$ compared to the true value $\mathcal Q_{true}$.
The MSE of the predicted value of the quantity of interest $\mathcal Q^{(i)}$ with respect to the true value $\mathcal Q_{true}$ was computed as
\begin{equation}
    \text{MSE} = \frac{1}{N^\dagger} \sum_{i=1}^{N^\dagger} (\mathcal Q_{true} - \mathcal Q^{(i)})^2
\end{equation}
Similarly, the sample variance of $\mathcal Q$ for each experiment was computed as:
\begin{equation}
    s^2 = \frac{1}{N^\dagger - 1} \sum_{i=1}^{N^\dagger} (\mathcal Q^{(i)} - \bar{\mathcal Q})^2
\end{equation}
Finally, we constructed Gaussian kernel posterior density estimates $\hat{f}_{\pi(\theta|\mathbf{d})}(\mathcal{Q})$ from the posterior samples from each experiment $\{\mathcal Q^{(i)}\}_{i=1}^{N^\dagger}$, and computed the kernel density of the true value $\mathcal Q_{true}$ with respect to this density estimate. Kernel density estimates were computed using \texttt{SciPy} \citep{2020SciPy-NMeth} with automatic bandwidth determination \citep{scott_multivariate_1992}.

\subsubsection{Results}
We compared the MSE, variance, and kernel density of both the vanilla and dual-weighted posterior samples with the corresponding values for the initial posterior samples for all $n=30$ experiments. 

\begin{figure}[htbp]
    \begin{subfigure}{0.5\textwidth}
        \centering
        \includegraphics[width=\textwidth]{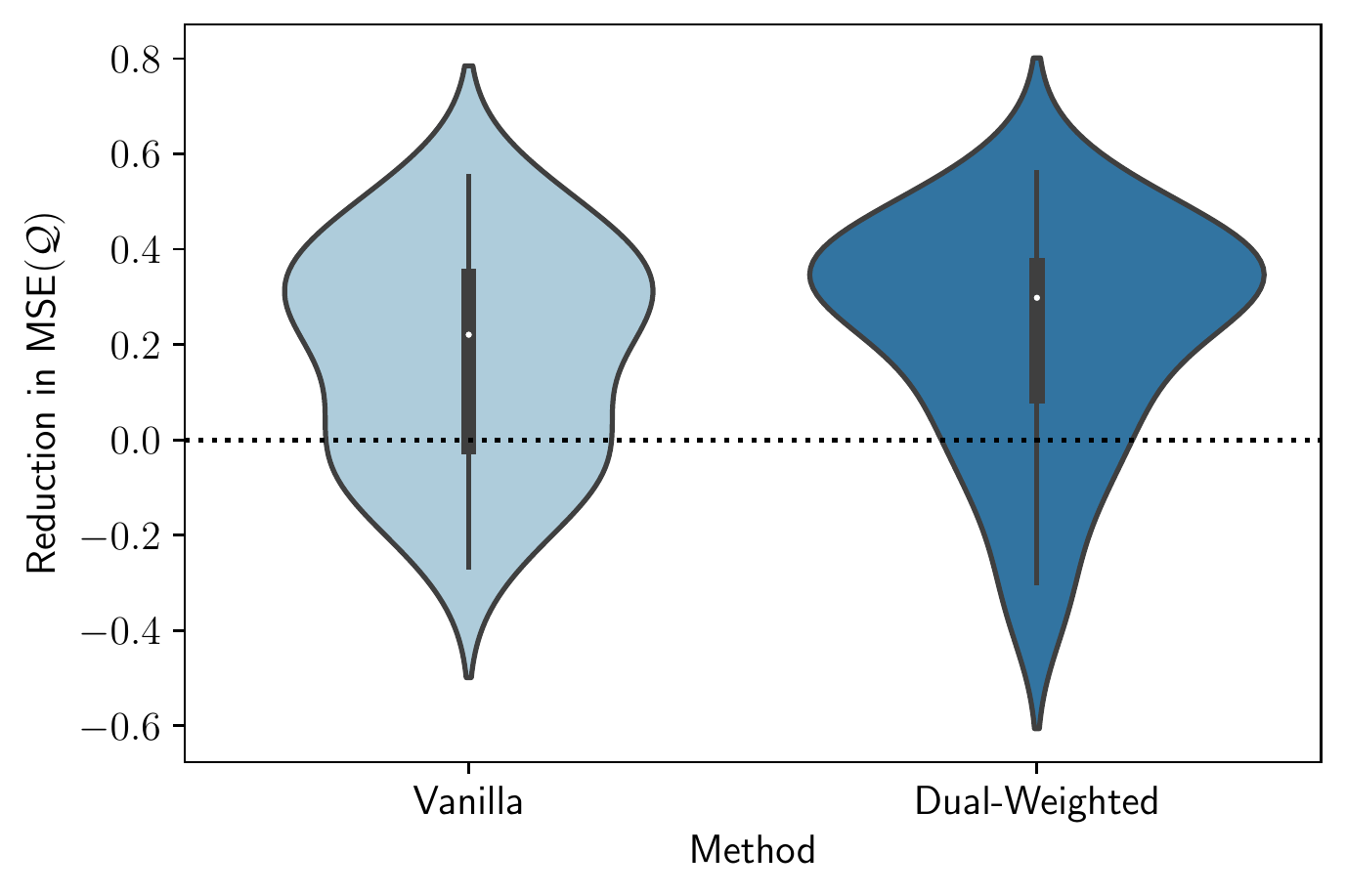}
        \caption{Reduction in MSE$(\mathcal{Q})$}
        \label{fig:MSE_violin}
    \end{subfigure}
    \begin{subfigure}{0.5\textwidth}
        \centering
        \includegraphics[width=\textwidth]{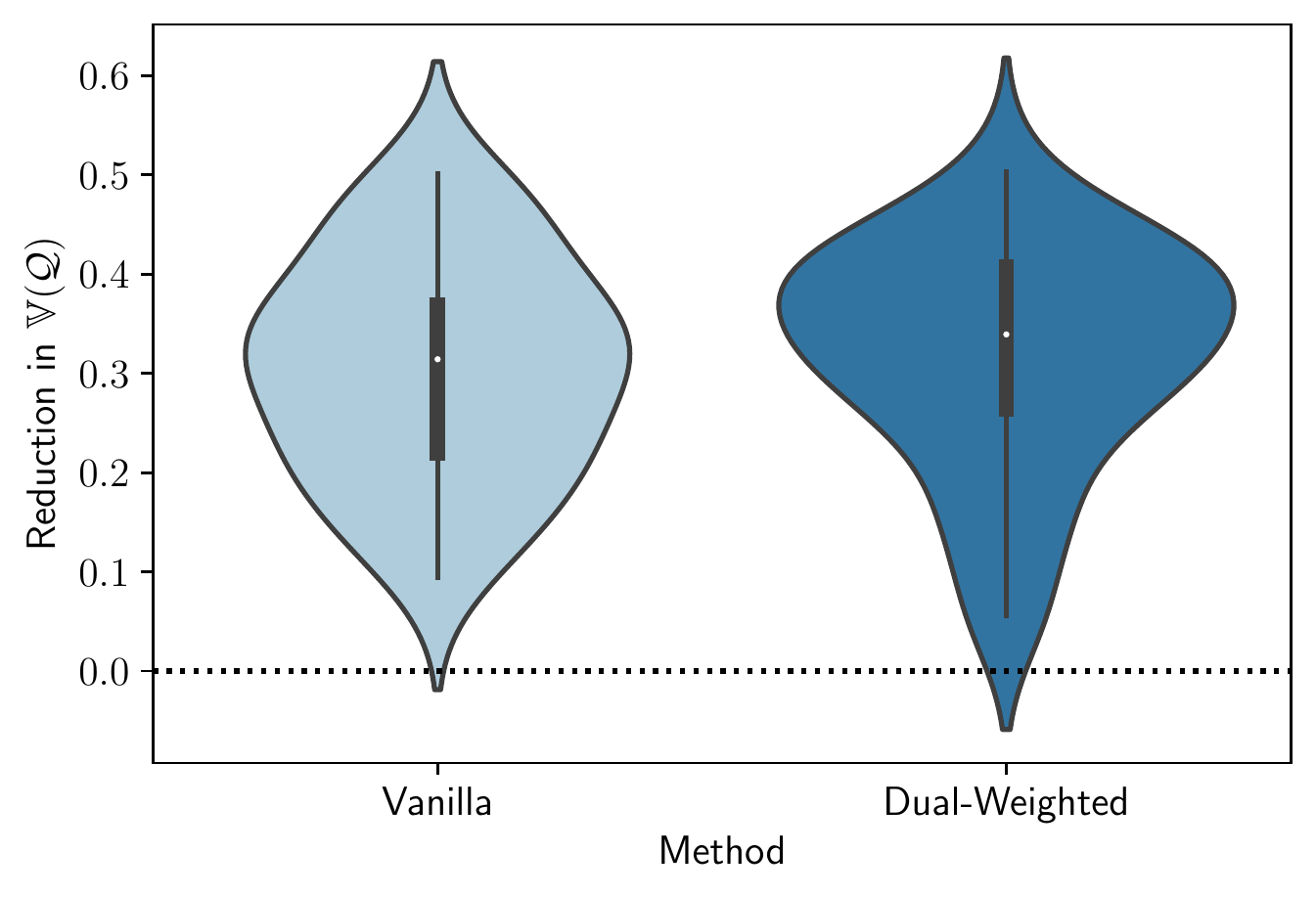}
        \caption{Reduction in $\mathbb{V}(\mathcal{Q})$}
        \label{fig:variance_violin}
    \end{subfigure}
  \caption{Kernel densities of the sample error of the quantity of interest $\varepsilon^{(i)} = Q_{true} - Q^{(i)}$ for the initial, vanilla and dual-weighted posteriors for two samples of the $n=30$ experiments.}
  \label{fig:violin}
\end{figure}

With respect to the MSE, the vanilla approach yielded a median reduction of $22\%$, while the dual--weighted approach yielded a median reduction of $30\%$ (Fig. \ref{fig:MSE_violin}). This demonstrates that both acquisition strategies approach the true value when we add more datapoints, but that the dual-weighted approach is more efficient. With respect to the variance of the quantity of interest, the vanilla approach yielded a median reduction of $31\%$, while the dual--weighted approach yielded a median reduction of $34\%$ (Fig. \ref{fig:variance_violin}). This shows that for both acquisition strategies the posterior distribution contracts as more data is added, and that the two approaches differ less with respect to this feature. However, this metric shows only that the posterior contracts, and not if it moves closer to the true value. Finally, we computed the posterior densities of the true quantity of interest with respect to kernel posterior density estimates $\hat{f}_{\pi(\theta|\mathbf{d})}(\mathcal{Q})$ for each experiment. Here, the vanilla approach yielded a median improvement of $12\%$, while the dual--weighted approach yielded a median improvement of $17\%$. Since the prediction variance of the quantity of interest reduced in every experiment (Fig. \ref{fig:variance_violin}), this again shows that the posterior distribution moves closer to the true value as more data is added, but that the dual-weighted approach is better.

We note that in neither method was capable of improving the posterior estimate of the quantity of interest for every experiment. Hence, in $8/30$ vanilla experiments and $5/30$ dual-weighted experiments, adding additional wells resulted in a worse posterior MSE than the initial one. This is not surprising since we are dealing with a very ill-posed inverse problem, and any new datapoint may reinforce the initial bias rather than reduce it. While both approaches occasionally failed to improve the posterior estimate, the dual-weighted approach performed better than the vanilla approach.

We computed the Gaussian kernel density estimates of the error $\varepsilon^{(i)} = \mathcal Q_{true} - \mathcal Q^{(i)}$ for two samples of the $n=30$ experiments. The left panel shows a typical example, where the vanilla approach resulted in a moderate improvement while the dual-weighted approach yielded a more dramatic improvement. The right panel shows an example where both the dual-weighted and vanilla approaches failed to produce any improvement.

\begin{figure}[htbp]
    \begin{subfigure}{0.5\textwidth}
        \centering
        \includegraphics[width=0.9\textwidth]{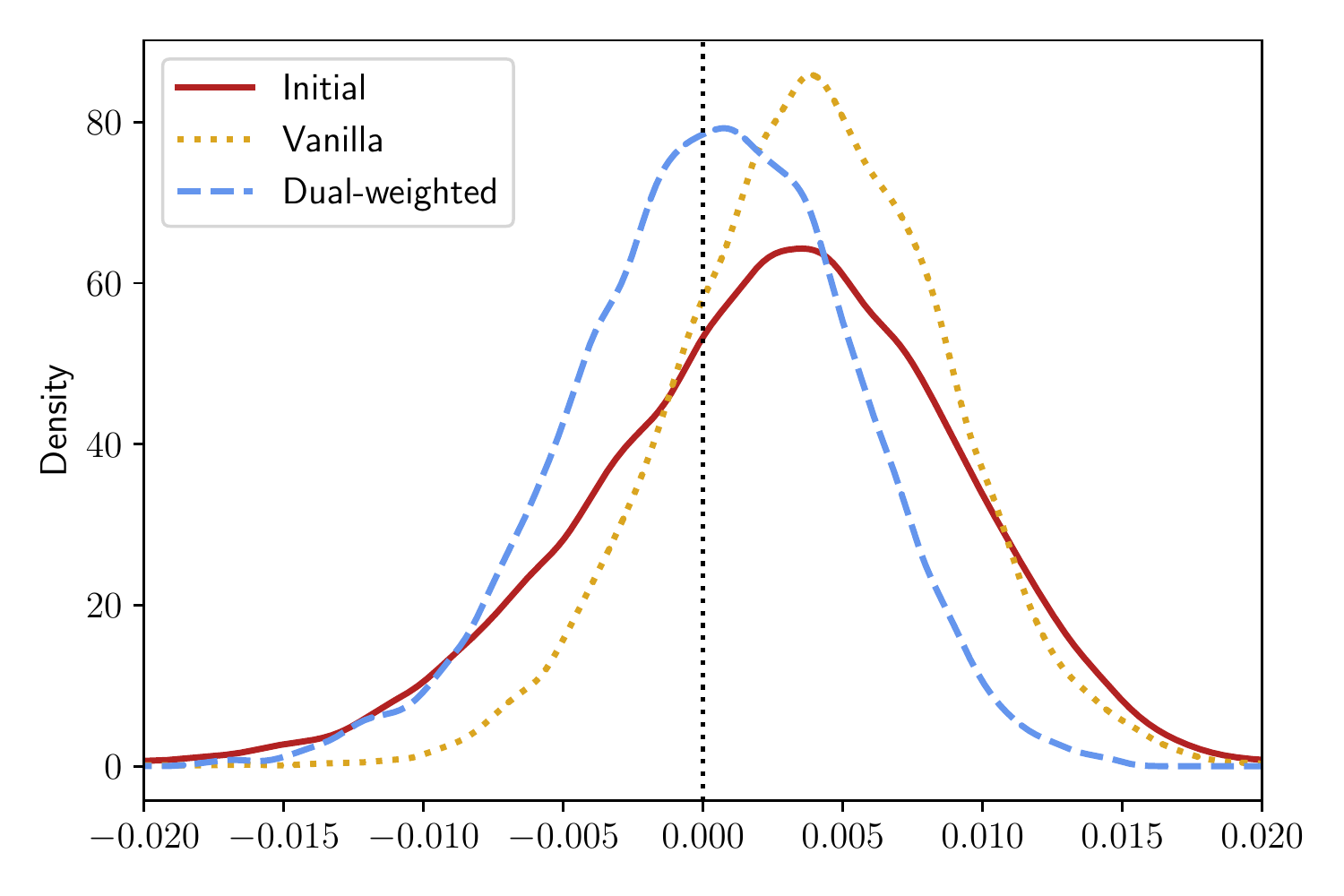}
        \caption{}
        \label{fig:density_10}
    \end{subfigure}
    \begin{subfigure}{0.5\textwidth}
        \centering
        \includegraphics[width=0.9\textwidth]{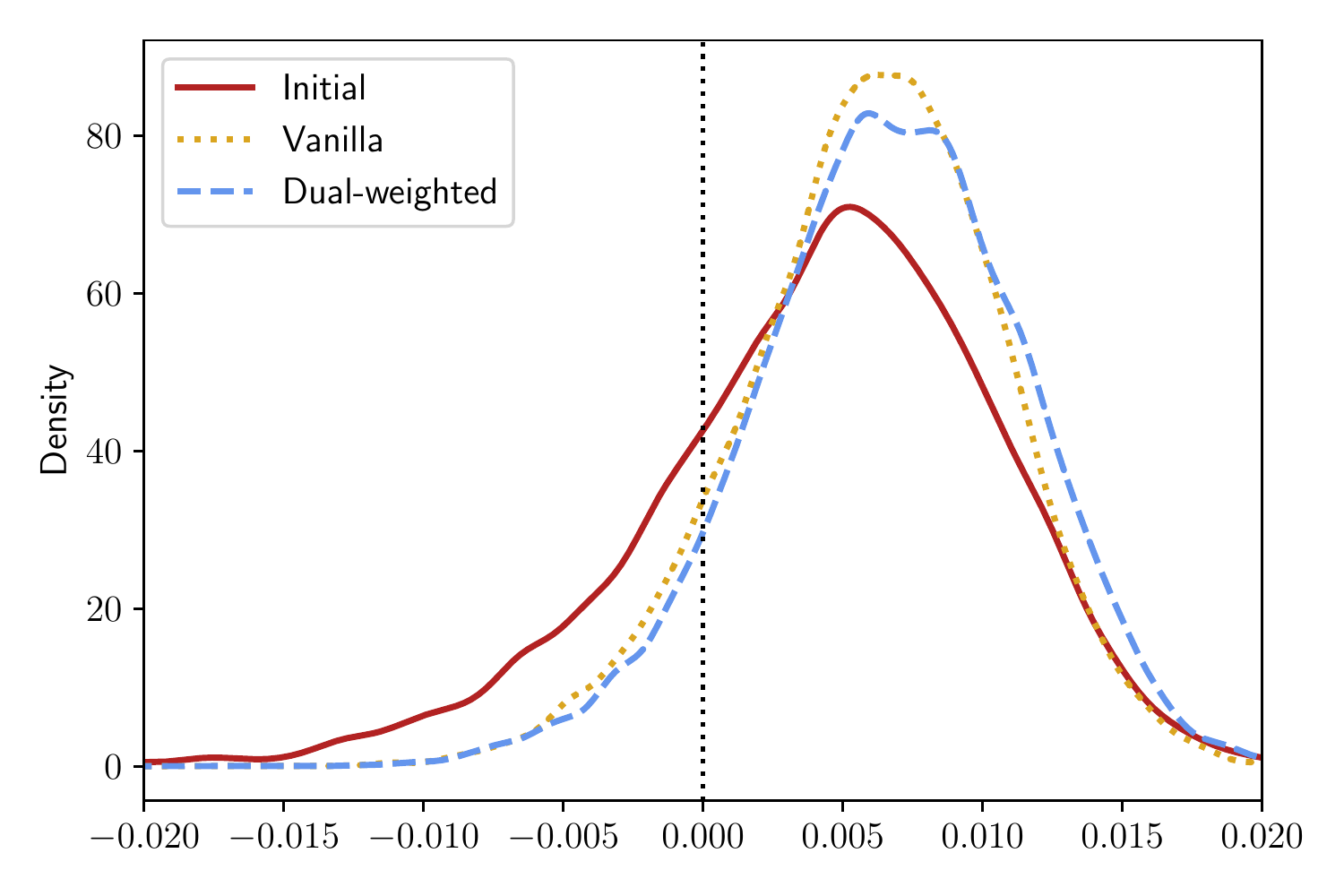}
        \caption{}
        \label{fig:density_28}
    \end{subfigure}
  \caption{Kernel densities of the sample error of the quantity of interest $\varepsilon^{(i)} = Q_{true} - Q^{(i)}$ for the initial, vanilla and dual-weighted posteriors for two samples of the $n=30$ experiments.}
  \label{fig:density}
\end{figure}

\section{Discussion} \label{sec:discussion}
In this paper, we have proposed a novel approach to the problem of optimally choosing the next location for a monitoring well, given existing data and some quantity of interest (QoI). The proposed methodology exploits the solution of an adjoint problem to weigh such an acquisition function according to the expected influence on the QoI. Numerical experiments have demonstrated that the approach works for our model problem. We emphasize that the problem is intrinsically probabilistic, and hence subject to uncertainty. We have demonstrated that the approach works \textit{on average} for our model problem, but there were certain experiments, where the dual-weighted acquisition strategy did not approach the true QoI (see e.g. Fig. \ref{fig:density_28}). As the number of wells approach infinity, the posterior distribution will certainly approach the true value, but for any one new observation well, there are no such guarantees. In a sense, the dual-weighted approach merely increases the chance of improving the posterior distribution of the QoI.

While we formulated and demonstrated the approach in the context of a groundwater surveying problem, the method could be applicable to other areas of science and engineering, where measurements are expensive. The most obvious parallel application is petroleum engineering, where there are similarities both in terms of the constituent equations and the mode of sampling, but the method could be adapted with little effort to any inverse problem where establishing sensors is expensive. We note, however, that the dual problem in our case was unusually simple, since the groundwater flow equation is self-adjoint. Clearly, the dual-weighted approach can only be used as-written for QoIs, where an adjoint problem can be formulated and solved directly. For more complicated QoIs, an alternative approach would be to perturb the posterior mean or mode to approximate the influence function. Using such an approach would yield $\omega(\mathbf{x}, \mathbb E[\theta])$ rather than $\mathbb E[\omega(\mathbf{x},\theta)]$ as a weighting function.

A bottleneck of our approach is that the MCMC sampler is rerun after each (batch) data acquisition. Running MCMC for expensive forward models is notoriously computationally demanding, and while we employ various tricks to reduce the cost (such as Delayed Acceptance and proposal adaptivity), this is not the most elegant approach. One way to significantly alleviate the cost of subsequent posterior distributions would be to employ a particle filter to sequentially reweigh MCMC samples according to the new data \citep{chopin_sequential_2002}. This sequential approach was investigated in this study but it did not work well, mainly because of very high sample degeneracy. When the variance of the solution, as in our case, is relatively high at unobserved locations, only few posterior samples fit the new observations well, with the mentioned sample degeneracy as a result. Moreover, we found that the dispersion measures in Eqs.~(\ref{eq:vanilla}), (\ref{eq:dual_weighted}), (\ref{eq:dual_weighted_batch}) and (\ref{eq:vanilla_batch}) where highly sensitive to this sample degeneracy. This challenge could be alleviated by drawing more posterior samples for the initial MCMC, but that would only offset the cost. We remark that this approach might work better for lower-dimensional problems than the one investigated in this study. We highlight this problem as a potential target for future research.

The methodology was demonstrated empirically in the context of a synthetic groundwater flow example. This gives rise to at least three additional interesting directions of future research. First, showing theoretically that the distribution of the quantity of interest does indeed converge faster to the true value when using the dual-weighted approach, and examining the mechanisms that govern this process in detail. Second, testing the method in practice in the context of an actual groundwater survey. While testing the method in practice would certainly expose limitations and complications that were not identified in this study, it would be difficult to validate the method further in this fashion, since the true value of the QoI is rarely known in reality. This may be overcome by testing the method under controlled (laboratory) conditions. Third, generalising the dual-weighted approach to a wider range of PDE problems with different constituent equations and QoIs.

\section*{Acknowledgements}
The MCMC code used for Delayed Acceptance sampling can be found at \href{https://github.com/mikkelbue/tinyDA}{https://github.com/mikkelbue/tinyDA}, and additional code will be made available in the Open Research Exeter data repository upon publication at \href{https://ore.exeter.ac.uk/repository/}{https://ore.exeter.ac.uk/repository/}.
ML was funded as part of the Water Informatics Science and Engineering Centre for Doctoral Training (WISE CDT) under a grant from the Engineering and Physical Sciences Research Council (EPSRC), grant number EP/L016214/1. TD was funded by a Turing AI Fellowship (2TAFFP\textbackslash100007).
The authors would like to thank Robert Scheichl and Karina Koval for advice with regards to the formulation of the adjoint state equation.
The authors have no competing interests.

\begin{appendices}

\section{Adjoint State Equations} \label{ap:adjoint_equation}
\subsection{Domain Integral as Objective Function}
Given an objective function defined as an integral over the entire domain
\begin{equation}
    \mathcal Q = \int_\Omega f \: dx
\end{equation}
\citet[Eq.~(15)]{sykes_sensitivity_1985} write the derivative of $\mathcal Q$ with respect to some parameter $\alpha$ as
\begin{align}
\begin{split}
    \frac{d\mathcal Q}{d\alpha} &= \int_\Omega \left[ \frac{\partial f}{\partial \alpha} + \psi \left(\frac{\partial f}{\partial u} + \nabla \cdot k \nabla \omega \right) + \omega \frac{\partial g}{\partial \alpha} - \nabla \omega \cdot \frac{\partial k}{\partial \alpha} \nabla u \right] dx \\
                       &+ \int_\Gamma \left[ \psi (k \nabla \omega) \cdot \mathbf{n} + \omega \frac{\partial q_N}{\partial \alpha}  \right] ds
\end{split}
\end{align}
To eliminate the unknown state sensitivities $\psi = \frac{\partial u}{\partial \alpha}$ they solve
\begin{equation}
    \nabla \cdot k \nabla \omega + \frac{\partial f}{\partial u} = 0
\end{equation}
with boundary conditions $\omega_{D} = 0$ on $\Gamma_D$ and $q_N^\omega = k\nabla \omega \cdot \mathbf{n} = 0$ on $\Gamma_N$.

\subsection{Boundary Integral as Objective Function}
The problem addressed in this paper involves an objective function defined on a fixed-head boundary $\Gamma'$:
\begin{equation}
    \mathcal Q = \int_{\Gamma'} f \: ds \quad \text{with} \quad f = q = -k\nabla u \cdot \mathbf{n}^+
\end{equation}
Where $\mathbf{n}^+$ is the outward normal. Hence, the derivative of the objective function instead takes the form
\begin{align}
\begin{split}
    \frac{d \mathcal Q}{d\alpha} &= \int_\Omega \left[ \psi \left(\nabla \cdot k \nabla \omega \right) + \omega \frac{\partial g}{\partial \alpha} - \nabla \omega \cdot \frac{\partial k}{\partial \alpha} \nabla u \right] dx \\
                       &+ \int_\Gamma \left[ \psi (k \nabla \omega) \cdot \mathbf{n}^- + \omega \left( \frac{\partial \mathbf{q}}{\partial \alpha} \cdot \mathbf{n}^- + \frac{\partial \mathbf{q}}{\partial u} \psi \cdot \mathbf{n}^- \right) \right] ds \\
                       &+ \int_{\Gamma'} \left[ \frac{\partial f}{\partial \alpha} + \frac{\partial f}{\partial u} \psi \right] ds
\end{split}
\end{align}
where $\mathbf{n}^-$ is the inward normal \citep{sykes_sensitivity_1985} and
\begin{equation}
    \frac{\partial \mathbf{q}}{\partial \alpha} \cdot \mathbf{n}^- +\frac{\partial \mathbf{q}}{\partial u} \psi \cdot \mathbf{n}^- = \frac{\partial q_N}{\partial \alpha} \quad \text{on} \quad \Gamma_N.
\end{equation}
To eliminate the unknown state sensitivities $\psi$, we now solve 
\begin{equation}
    \nabla \cdot k \nabla \omega = 0
\end{equation}
with boundary conditions $\omega_{D} = 0$ on $\Gamma_D\setminus\Gamma'$ and $q_N^\omega = k\nabla \omega \cdot \mathbf{n}^- = 0$ on $\Gamma_N$. For the remaining boundary $\Gamma'$, we impose
\begin{equation} \label{eq:gamma_bc}
    \frac{\partial f}{\partial u} + \omega \frac{\partial \mathbf{q}}{\partial u} \cdot \mathbf{n}^- = 0.
\end{equation}
Since on $\Gamma'$ we have
\begin{equation} \label{eq:gamma_equality}
    - \frac{\partial \mathbf{q}}{\partial u} \cdot \mathbf{n}^- = \frac{\partial f}{\partial u}
\end{equation}
we can substitute (\ref{eq:gamma_equality}) into (\ref{eq:gamma_bc}) to get 
\begin{equation}
    \frac{\partial f}{\partial u} - \omega \frac{\partial f}{\partial u} = 0 \quad \text{on} \quad \Gamma'
\end{equation}
and so the operative boundary condition on $\Gamma'$ is $\omega_{\Gamma'} = 1$.

\end{appendices}

\bibliographystyle{unsrtnat}  
\bibliography{main} 

\end{document}